\documentclass[pdftex, noinfoline, letter]{imsart}
 
\usepackage[english]{babel}
\usepackage{parskip}
\usepackage{amsmath,amsfonts,amsthm, amssymb}
\usepackage{graphicx} 
\usepackage{enumitem} 
\usepackage{xcolor, latexsym} 
\usepackage{subfigure} 
\usepackage{hyperref} 
\usepackage[normalem]{ulem} 
\usepackage{float}
\usepackage{placeins}
\usepackage{soul} 
\soulregister\cite7 
\soulregister\ref7
\soulregister\pageref7

\DeclareMathOperator{\Tr}{Tr}

\DeclareMathOperator{\MF}{\text{MF}}

\DeclareMathOperator*{\argmin}{arg\,min}
\DeclareMathOperator*{\argmax}{arg\,max}

\DeclareMathOperator*{\bgamma}{\boldsymbol{\gamma}}
\DeclareMathOperator*{\ELBO}{\text{ELBO}}

\definecolor{blendedblue}{rgb}{0.2,0.2,0.7}
\definecolor{darkpurple}{RGB}{49,0,94}
\definecolor{darkgreen}{RGB}{11, 84, 37}
\newcommand{\widesim}[2][2]{
	\mathrel{\overset{#2}{\scalebox{#1}[1]{$\sim$}}}
}

\allowdisplaybreaks

\theoremstyle{plain} 
\newtheorem{theorem}{Theorem}[section]

\newtheorem{lemma}[theorem]{Lemma}
\newtheorem{assumption}{Assumption}

\allowdisplaybreaks

\endlocaldefs
\usepackage{natbib}
\usepackage{multirow}
\usepackage{chngpage}
\usepackage{siunitx}
\usepackage{booktabs}
\usepackage{makecell}
\usepackage[T1]{fontenc}
\usepackage{float}
\usepackage{grffile}
\usepackage{url} 
\usepackage{graphicx,psfrag,epsf}

\usepackage{bbm, dsfont}
\usepackage{mathrsfs} 
\usepackage[ruled,vlined]{algorithm2e}
\SetKwInput{KwInput}{Input}                
\SetKwInput{KwOutput}{Output}
\allowdisplaybreaks

\begin{document}

\begin{frontmatter}

\title{Spike and slab Bayesian sparse principal component analysis}

\runtitle{Spike and slab Bayesian SPCA}

\author{\fnms{Bo Y.-C.} \snm{Ning}\ead[label=e1]{yuchien.ning@gmail.com}}
\and
\author{\fnms{Ning} \snm{Ning}\ead[label=e2]{patning@tamu.edu}}

\affiliation{Harvard University\thanksmark{m1}}

\address{Harvard T. H. Chan School of Public Health \\ 
Department of Epidemiology, \\
677 Huntington Ave, Boston, MA 02115\\
  \printead{e1}}
\address{Department of Statistics\\
	Texas A\&M University\\
	College Station, TX 77843
	\printead{e2}}

\runauthor{Ning and Ning}

\begin{abstract}
 Sparse principal component analysis (SPCA) is a popular tool for dimensionality reduction in high-dimensional data. However, there is still a lack of theoretically justified Bayesian SPCA methods that can scale well computationally. One of the major challenges in Bayesian SPCA is selecting an appropriate prior for the loadings matrix, considering that principal components are mutually orthogonal. We propose a novel parameter-expanded coordinate ascent variational inference (PX-CAVI) algorithm. This algorithm utilizes a spike and slab prior, which incorporates parameter expansion to cope with the orthogonality constraint. Besides comparing to two popular SPCA approaches, we introduce the PX-EM algorithm as an EM analogue to the PX-CAVI algorithm for comparison. Through extensive numerical simulations, we demonstrate that the PX-CAVI algorithm outperforms these SPCA approaches, showcasing its superiority in terms of performance. We study the posterior contraction rate of the variational posterior, providing a novel contribution to the existing literature. The PX-CAVI algorithm is then applied to study a lung cancer gene expression dataset. The $\mathsf{R}$ package $\mathsf{VBsparsePCA}$ with an implementation of the algorithm is available on the Comprehensive R Archive Network (CRAN). 
\end{abstract}


\end{frontmatter}


\section{Introduction}
\label{sec:intro}

Sparse Principal Component Analysis (SPCA), a contemporary variant of PCA, has gained popularity as a valuable tool for reducing the dimensions of high-dimensional data. Its applications span various fields, such as chemistry, where it aids in identifying crucial chemical components from spectra \citep{varmuza09}; genetics, where it helps discover significant genes and pathways \citep{li17}; and macroeconomics, where it plays a role in selecting dominant macro variables that earn substantial risk premiums \citep{rapach19}. The success of SPCA can be attributed to two main factors. Firstly, in typical high-dimensional datasets, the number of input variables $p$ is greater than the number of observations $n$. This condition poses challenges when using traditional PCA, as the leading eigenvector becomes inconsistently estimated when $p/n$ does not converge to 0 \citep{paul07, johnstone09}. However, SPCA addresses and mitigates this issue effectively. Secondly, the principal components derived from SPCA are linear combinations of only a few important variables, making them highly interpretable in practical applications. This interpretability makes SPCA a valuable asset when dealing with complex data sets, enabling researchers and analysts to glean meaningful insights with ease.

Several SPCA algorithms have been proposed, and interested readers can refer to \citet{zou18} for a comprehensive literature review on these algorithms. However, it's worth noting that the algorithms discussed in that review do not include Bayesian-based methods. Recently, two Bayesian SPCA approaches, introduced by \citet{gao15} and \citet{xie18}, have emerged and demonstrated impressive advantages. Both approaches adopt the spiked covariance model, which conveniently represents a linear regression model where the loadings matrix serves as the coefficient and the design matrix follows a standard multivariate normal distribution as the random component. A significant challenge in Bayesian SPCA lies in placing a prior on the coefficients while enforcing the orthogonality constraint, which requires the columns of the loadings matrix to be mutually orthogonal. This constraint needs to be incorporated through the prior distribution. \citet{gao15} tackled this challenge by constructing a prior that projects the nonzero coordinates onto a subspace spanned by a collection of mutually orthogonal unit vectors. However, their posterior becomes intractable and challenging to compute when the rank is greater than one.
\citet{xie18} adopted a different approach by reparametrizing the likelihood, multiplying the loadings matrix with an orthogonal matrix to remove the orthogonal constraint. Their prior involves only Laplacian spike and slab densities while our density (see Equation \eqref{prior-beta} in Section \ref{sec:prior}) is considerably more general. Additionally, this prior can introduce dependence while theirs demand prior independence. 

We present a novel prior for the coefficient of the spiked covariance model. In our approach, we apply a regular spike and slab prior on the parameter, which is the product of the loadings matrix (the coefficient) and an orthogonal matrix. The orthogonal matrix is then a latent variable in the prior. By marginalizing this joint density, we derive the prior of the coefficient. The spike and slab prior is a mixture of a continuous density and a Dirac measure centered at 0. By introducing an appropriate prior on the mixture weight, one can effectively impose sparsity on the coefficient. The spike and slab prior is widely recognized as one of the most prominent priors for Bayesian high-dimensional analysis and has received extensive study. Excellent works in this area include those by \citet{johnstone04, rockova18a,cast20a,cast12, cast15, martin17,  qiu2018multivariate, jammalamadaka2019predicting, qiu2020multivariate, jeong20, ohn23, ning23}, and \citet{ning20}. For a comprehensive overview of this topic, readers can refer to the review paper by \citet{banerjee20}.
It is important to note that in our spike and slab formulation, the slab density, which incorporates the latent variable, differs from that in traditional linear regression models. This distinction contributes to the uniqueness and effectiveness of our proposed approach.

We employ a variational approach to compute the posterior, a method that minimizes a chosen distance or divergence (e.g., Kullback-Leibler divergence) between a preselected probability measure, belonging to a rich and analytically tractable class of distributions, and the posterior distribution. This approach offers faster computational speed compared to sampling methods like the Markov chain Monte Carlo algorithm. Among the variational approaches, the coordinate ascent variational inference (CAVI) method stands out as the most popular algorithm \citep{blei17}.
 Several CAVI methods have been developed for sparse linear regression models with the spike and slab prior (or the subset selection prior) such as \cite{carbonetto12, huang16, ray20, yang20}. Researchers have also studied the theoretical properties of the variational posterior, such as the posterior contraction rate, as examined by \citet{ray20} and \citet{yang20}.
While variational Bayesian methods for SPCA have been developed by \citet{guan09} and \citet{bouveyron18}, they did not provide a theoretical justification for their posterior. Moreover, the priors used by \citet{guan09} involving the Laplace distribution and \citet{bouveyron18}'s prior, similar to the spike and slab prior with a fixed mixture weight, are known not to yield the optimal (or near-optimal) posterior contraction rate.

In this paper, we show that the contraction rates of both the posterior and the variational posterior are nearly optimal. To the best of our knowledge, this is the first result for the variational Bayesian method applied to SPCA.
Additionally, we develop an EM algorithm tailored for SPCA,  in which the maximum of a posteriori estimator is obtained. The EM algorithm for Bayesian variable selection has been extensively studied for the sparse linear regression model by \citet{rockova14b,rockova18a}. Similar algorithms have been developed for other high-dimensional models, such as the dynamic time series model \citep{ning19} and the sparse factor model \citep{rockova16}. For our EM algorithm to accommodate SPCA, we replace the Dirac measure in the spike and slab prior with a continuous density, resulting in the continuous spike and slab prior.
Both the variational approach and the EM algorithm employ parameter expansion techniques on the likelihood function. Consequently, these algorithms are referred to as the PX-CAVI and the PX-EM algorithm respectively, where PX means parameter expanded. The parameter expansion approach was initially proposed by \citet{liu98} and has proven effective in accelerating the convergence speed of the EM algorithm. Additionally, we discovered that by selecting the expanded parameter as the orthogonal matrix, we can circumvent the need to handle the orthogonal constraint directly on the loading matrix. This approach allows us to first solve for the unconstrained matrix and subsequently apply singular value decomposition (SVD) to obtain an estimated value for the loadings matrix. This simplification streamlines the computation process and enhances the efficiency of our algorithms.

The remainder of this paper is structured as follows:
Section \ref{sec:model-and-priors} presents the model and the prior used in this study.
Section \ref{sec:VI} introduces the variational approach and outlines the development of the PX-CAVI algorithm.
In Section \ref{sec:postCR}, we delve into the theoretical properties of both the posterior and the variational posterior.
Section \ref{sec:EM} presents the PX-EM algorithm we developed.
To evaluate the performance of our algorithms, we conduct simulation studies in Section \ref{sec:simstudy}.
Furthermore, in Section \ref{sec:realstudy}, we analyze a lung cancer gene dataset to illustrate the application of our approach in real-world scenarios.
The appendix contains proofs of the equations presented in Section \ref{sec:VI}.
Proofs of the theorems discussed in Section \ref{sec:postCR} and the batch PX-CAVI algorithm without relying on the jointly row-sparsity assumption are provided in the supplementary material. For readers interested in implementing our algorithms, we have made the $\mathsf{VBsparsePCA}$ package available on the comprehensive R archive network (CRAN). This package includes both the PX-CAVI algorithm and the batch PX-CAVI algorithm.

\section{Model and priors}
\label{sec:model-and-priors}

In this section, we begin by introducing the spiked covariance model, followed by the spike and slab prior applied.

\subsection{The spiked covariance model}

Consider the spiked covariance model 
\begin{align}
X_i = \theta w_i + \sigma \epsilon_i, \quad
w_i \widesim{\text{i.i.d.}} \mathcal{N}(0, I_r), \quad
\epsilon_i \widesim{\text{i.i.d.}} \mathcal{N}(0, I_p),
\label{model}
\end{align}
where $X_i$ is a $p$-dimensional vector, $\theta$ is a $p \times r$-dimensional loadings matrix, 
$w_i$ is a $r$-dimensional vector, 
$\epsilon_i$ is a $p$-dimensional vector that is independent of $w_i$, and $r$ is the rank. We denote $\theta_{\cdot k}$ as the $k$-th column of $\theta$. 
The orthogonality constraint of $\theta$ requires that $\langle \theta_{\cdot k}, \theta_{\cdot k'} \rangle = 0$ for any $k \neq k'$, $k, k' \in \{1, \dots, r\}$. 
The model is equivalent to $X_i \widesim{\text{i.i.d.}} \mathcal{N}(0, \Sigma)$, where $\Sigma = \theta \theta' + \sigma^2 I_p$. 
Let $\theta = U\Lambda^{1/2}$, where 
$U = \left({\theta_{\cdot 1}}/{\|\theta_{\cdot 1}\|_2}, \dots, {\theta_{\cdot r}}/{\|\theta_{\cdot r}\|_2}\right)$ is a $p \times r$ matrix containing the first $r$ eigenvectors 
and $\Lambda = \text{diag}(\|\theta_{\cdot 1}\|_2^2, \dots, \|\theta_{\cdot r}\|_2^2)$ is an $r \times r$ diagonal matrix.
Then, $\Sigma = U\Lambda U' + \sigma^2 I_p$. One can easily check that the $k$-th eigenvalue of $\Sigma$ is $\|\theta_{\cdot k}\|_2^2 + \sigma^2$ if $k \leq r$ and is $\sigma^2$ if $k > r$.
We assume $p \gg n$ (i.e. $n/p \to 0$) and $\theta$ is jointly row-sparse---that is, within the same row, the coordinates are either all zero or all non-zero. 
We define the rows containing non-zero entries as ``non-zero rows'' and the remaining rows as ``zero rows.'' With this assumption, the support of each column in $\theta$ remains the same and is denoted as 
$S = \left\{j \in \{1, \dots, p\}: \ \theta_{j}' \neq 0_r \right\}$
where $0_r$ represents $r$-dimensional zero vector.
Adopting the row-sparsity assumption is convenient for practitioners, as the principal subspace is generated by the same sparse set of features. Additionally, we can simplify our main ideas and use more concise notations by adopting this assumption, as the support is consistent across all principal components. A more general assumption that allows the support to vary across principal components, is covered in the supplementary material. Our $\mathsf{R}$ package $\mathsf{VBsparsePCA}$ can effectively handle both assumptions.

\subsection{The spike and slab prior}
\label{sec:prior}

We introduce our spike and slab prior, which is 
\begin{align}
\label{spike-and-slab-prior}
& \pi(\theta, \bgamma|\lambda_1, r) \propto 
\prod_{j=1}^p 
\left[
	\gamma_j \int_{A \in V_{r,r}} g(\theta_j|\lambda_1, A, r) \pi(A) d A+ (1-\gamma_j) \delta_0(\theta_j)
\right],
\end{align}
where 
$V_{r,r} = \{
A \in \mathbb{R}^{r \times r}: A'A = I_r
\}$ is the Stiefel manifold of $r$-frames in $\mathbb{R}^r$ and $\delta_0$ is the Dirac measure at zero.
Our idea of constructing the prior (\ref{spike-and-slab-prior}) is that since $\beta = \theta A$ does not have the orthogonality constraint, as $A$ is an orthogonal matrix, we first apply the regular spike and slab prior on $\beta$ which could be viewed as the joint distribution of $\theta$ and $A$. We then obtain the prior of $\theta$ by marginalizing the parameter $A$ from the joint distribution of $\theta$ and $A$.
Because of the latent variable $A$, this prior is different from those in the sparse linear regression models.
We consider a general expression for the density $g$, which is
\begin{align}
\label{prior-beta}
g(\theta_j|\lambda_1, A, r) = [C(\lambda_1)]^r \exp(-\lambda_1 \|\beta_j\|_q^m),
\end{align}
where $1 \leq q \leq 2$, $m \in \{1, 2\}$, and $[C(\lambda_1)]^r$ is the normalizing constant.
This expression includes three common distributions as special cases. If $q = 1$ and $m = 1$, $C(\lambda_1) = \lambda_1/2$, then $g$ is a product of $r$-independent Laplace densities. If $q = 2$ and $m = 2$, $C(\lambda_1) = \sqrt{\lambda_1/(2\pi)}$, then it is the multivariate normal density. If $q = 2$ and $m = 1$, $C(\lambda_1) = \lambda_1/a_r$ where $$a_r = \sqrt{\pi}\left(\Gamma(r+1)/\Gamma(r/2+1)\right)^{1/r} > 2,$$ then it is the density part of the prior introduced by \citet{ning20} for group sparsity.
The priors for the remaining parameters are given as follows: $\pi(A) \propto 1$ and for each $j$, 
\begin{align}
\label{prior-gamma}
& \gamma_j| \kappa \sim \text{Bernoulli}(\kappa), \qquad \kappa \sim \text{Beta}(\alpha_1, \alpha_2).
\end{align}
If $\sigma^2$ and $r$ are unknown, we let $\sigma^2 \sim \text{InverseGamma}(\sigma_a, \sigma_b)$ and $r \sim \text{Poisson}(\varkappa)$.
Assuming $r$ is fixed, then the joint posterior distribution of $(\theta, \bgamma, \sigma^2)$ is
\begin{align}
\label{posterior}
\pi(\theta, \bgamma, \sigma^2|X) 
\propto 
\prod_{i=1}^n f(X_i|\theta, \sigma^2, r) \prod_{j=1}^p \pi(\theta_j |\gamma_j, r) \left(\int \pi(\gamma_j|\kappa)\pi(\kappa)d\kappa\right) \pi(\sigma^2),
\end{align}
where $X=(X_1,\ldots,X_n)$ with each $X_i$ being a $p$-dimensional vector.
\section{Variational inference}
\label{sec:VI}

In this section, we propose a variational approach for SPCA using the posterior (\ref{posterior}). We introduce a mean-field variational class to obtain the variational posterior, and then develop the PX-CAVI algorithm to efficiently compute it.

\subsection{The variational posterior and the evidence lower bound}

To obtain the variational posterior, we adopt the mean-field variational approximation, which decomposes the posterior into several independent components, with the parameter in each component being independent of the others.
The variational class is defined as follows:
\begin{equation}
\begin{split}
\label{eqn:MF-class}
\mathcal{P}^{\text{MF}} = 
\Bigg\{
P(\theta):= 
& \prod_{j=1}^p \Big[
z_j \mathcal{N}(\mu_j, \sigma^2 M_j) + 
(1 - z_j) \delta_0
\Big],\; \mu_j \in \mathbb{R}^{r},\\
& \langle \mu_{\cdot k}, \mu_{\cdot k'} \rangle = 0, \;\forall k \neq k', \;
M_j \in \mathbb{M}^{r \times r},\;
z_j \in [0, 1]
\Bigg\},
\end{split}
\end{equation}
where $\mathbb{M}^{r \times r}$ stands for the space of $r \times r$ positive definite matrices. 
For any $P(\theta) \in \mathcal{P}^{\MF}$, it is a product of $p$ independent densities, each of which is a mixture of two distributions---a multivariate normal (or a normal density when $r = 1$) and the Dirac measure at zero. The mixture weight $z_j$ is the corresponding inclusion probability.
The variational posterior is obtained by minimizing the Kullback-Leibler divergence between all $P(\theta) \in \mathcal{P}^{\text{MF}}$ and the posterior, i.e.,
\begin{align}
\label{eqn:MF-posterior}
    \widehat  P(\theta) = \argmin_{P(\theta) \in \mathcal{P}^{\text{MF}}}
    KL\left(P(\theta), \pi(\theta|X)\right),
\end{align}
which can be also written as
\begin{align}
    \widehat P(\theta) 
    & = 
    \argmin_{P(\theta) \in \mathcal{P}^{\text{MF}}}
    \Big(\mathbb{E}_P \log P(\theta) - \mathbb{E}_P \log\pi(\theta|X)\Big) \nonumber \\
    & =
    \argmin_{P(\theta) \in \mathcal{P}^{\text{MF}}}
    \Big(
        \mathbb{E}_P \log P(\theta) - \mathbb{E}_P \log\pi(\theta, X)
        + \log \pi(X)
    \Big).
    \label{eqn:VP-1}
\end{align}
As the expression of $\log \pi(X)$ in \eqref{eqn:VP-1} is intractable, we define the evidence lower bound (ELBO), which is the lower bound of $\log \pi(X)$ as follows:
\begin{align}
\label{eqn:ELBO}
    \ELBO(\theta) & =
    \mathbb{E}_P \log\pi(\theta, X) - \mathbb{E}_P \log P(\theta).
\end{align}
and solve $\widehat  P(\theta_j) = \argmax_{P(\theta) \in \mathcal{P}^{\text{MF}}}\ELBO(\theta)$.
Since 
$$P(\theta) = \prod_{j=1}^p P(\theta_j)\quad \text{and} \quad\pi(\theta, X) = \prod_{j=1}^p \pi(\theta_j, X),$$ the ELBO can be also written as follows:
\begin{align*}
    \ELBO(\theta) = 
    \sum_{j=1}^p 
    \Big(
    \mathbb{E}_P \log\pi(\theta_j, X)
    -
    \mathbb{E}_P \log P(\theta_j)
    \Big).
\end{align*}
From the last display, we can solve each $\widehat  P(\theta_j)$ independently and then obtain
the variational posterior from $\widehat  P(\theta) = \prod_{j=1}^p \widehat  P(\theta_j)$.

\subsection{The PX-CAVI algorithm}
\label{sec:CAVI}

The PX-CAVI algorithm is an iterative method where, in each iteration, it optimizes each of the unknown variables by conditioning on the rest.
Our algorithm incorporates two key differences from the conventional CAVI algorithm. Firstly, we include an expectation step, similar to that used in the EM algorithm, since $w = (w_1, \dots, w_n)$ is a random variable. Secondly, we apply parameter expansion to the likelihood, which enables us to handle the orthogonality constraint and accelerate the convergence speed of our algorithm. Now, let's provide a step-by-step derivation of the PX-CAVI algorithm, where $M= (M_1, \dots, M_p)$ and 
$z = (z_1, \dots, z_p)$.
\medskip


{\bf 1. E-step} 

In this step, the full model posterior is $\pi(\theta, w, X)$. Let $\Theta^{(t)}$ be the estimated value of $\Theta = (\mu, M, z)$ from the $t$-th iteration, we obtain
\begin{eqnarray}
\label{eqn:cavi-w}
    & \widehat  P(w_i|\Theta^{(t)}) = \mathcal{N}(\widetilde  \omega_i, \widetilde  V_w), & \nonumber\\
    & \widetilde  \omega_i = \frac{1}{\sigma^{2}} \widetilde  V_w \sum_{j=1}^p z_j^{(t)} \big[\mu_j^{(t)}\big]' X_{ij}, &  \\
& \widetilde  V_w = 
\left(\frac{1}{\sigma^{2}}
\sum_{j=1}^p z_j^{(t)}
\left(\big[\mu_j^{(t)}\big]' {\mu_j^{(t)}} + \sigma^2M_j^{(t)} \right)
+ I_r
\right)^{-1}. &
\nonumber 
\end{eqnarray}
Then, the objective function is given by 
$$Q(\theta|\Theta^{(t)}) = \mathbb{E}_{w|\Theta^{(t)}}\log  \pi(\theta, w, X).$$
We obtain 
\begin{align*}
\widehat  P(\theta) & = 
\argmax_{P(\theta) \in \mathcal{P}^{\text{MF}}}
\sum_{j=1}^p
\left(
\mathbb{E}_P Q(\theta_j |\Theta^{(t)}) - \mathbb{E}_P \log P(\theta_j)
\right).
\\
& = \argmax_{P(\theta) \in \mathcal{P}^{\text{MF}}}
\sum_{j=1}^p
\Big(
\mathbb{E}_P \big[\mathbb{E}_{w|\Theta^{(t)}} \log \pi(\theta_j, w, X)\big] - \mathbb{E}_P \log P(\theta_j)
\Big).
\end{align*}

{\bf 2. Parameter expansion}

To obtain $\widehat P(\theta)$, special attention must be given to the orthogonality constraint of $\mu$ as defined in (\ref{eqn:MF-class}). This is where the parameter expansion technique is used. Let $A$ be the expanded parameter and denote $\beta = \theta A$, the likelihood after the parameter expansion becomes $X_i = \beta w_i + \sigma^2 \epsilon_i$, as $A w_i \widesim{\text{i.i.d}} \mathcal{N}(0, I_r)$ follows the same distribution as $w_i$. Then, our spike and slab prior is directly applied on $\beta$. We do not require the prior to be invariant under the transformation of the parameter. 
After solving $\beta$, one can obtain $\theta$ using the singular value decomposition (SVD). 
To accelerate the convergence speed of the algorithm, we apply parameter expansion again. At this time, the expanded parameter is chosen to be a positive definite matrix, say $D$. 
We denote $\widetilde  \beta = \beta D$. The likelihood after this parameter expansion becomes 
$X_i = \widetilde  \beta \widetilde  w_i + \sigma \epsilon_i$, where $\widetilde  w_i \widesim{\text{i.i.d}} \mathcal{N}(0, D)$ and $\widetilde  \beta = \widetilde  \theta A D_L^{-1}$, $D_L$ is the lower triangular matrix obtained using SVD. 
Our spike and slab prior is then directly putting on $\widetilde  \beta$. 
To summarize, parameter expansion is used twice in the PX-CAVI algorithm. The first time is primary used to deal with the orthogonality constraint, and the second time is to accelerate its convergence speed. We denote $\widetilde  u$ and $\widetilde  M$ as the mean and the covariance of $P(\widetilde  \beta)$. This leads us to  instead maximize $\mathbb{E}_P Q(\widetilde  \Theta|\widetilde  \Theta^{(t)}) - \mathbb{E}_{q} \log P(\widetilde  \beta)$, where 
$\widetilde  \Theta = (\widetilde  u, \widetilde  M, z)$ and 
\begin{align}
\label{eqn:tildebeta}
        P(\widetilde  \beta) = \prod_{j=1}^p 
        \left[
            z_j\mathcal{N}(\widetilde  u_j, \sigma^2 \widetilde  M_j) + (1-z_j) \delta_0
        \right].
\end{align}
One can quickly check that $\widetilde  u = \mu A D_L^{-1}$ and $\widetilde  M_j = D_L^{-1}  M_j {D_L^{-1}}'$. 
Note that since we assume $\theta$ is jointly row-sparse, the support of $\beta$ and it of $\widetilde  \beta$ are the same. Thus $z_j$ in (\ref{eqn:tildebeta}) is the same as it in (\ref{eqn:MF-class}).

To solve for $\widetilde u$ and $\widetilde M$, we explore the following two choices of the density $g$ in (\ref{prior-beta}):

$\bullet$ When $q = 1$ and $m = 1$, it yields a product of $r$-independent Laplace densities. Details are given in Appendix \ref{derive-cavi}. 
In summary, denoting $H_i = \widetilde  \omega_i \widetilde  \omega_i' + \widetilde  V_w$, we obtain 
\begin{align}
	\label{eqn:cavi-u}
	\widehat  {\widetilde  u}_j & = \min_{\widetilde  u_j}
	\left[
	\frac{1}{2\sigma^2} \sum_{i=1}^n
	\left(\widetilde  u_j H_i \widetilde  u_j' - 2 X_{ij} \widetilde  u_j \widetilde  \omega_i \right)
	+ \lambda_1 \sum_{k=1}^r f(\widetilde  u_{jk}, \sigma^2 \widetilde  M_{j,kk})
	\right] \\
	\label{eqn:cavi-Xi}
	\widehat  {\widetilde M}_j
	& = \min_{\widetilde  M_j}
	\left[\frac{1}{2} 
	\sum_{i=1}^n \Tr(\widetilde  M_j H_i)
	- \frac{\log \det(\widetilde  M_j) }{2}
	+ \lambda_1 \sum_{k=1}^r f(\widetilde  u_{jk},\sigma^2  \widetilde  M_{j, kk})
	\right],
\end{align}
where 
$f(\widetilde  u_{jk},\sigma^2  \widetilde  M_{j,kk})$ is the mean of the folded normal distribution, 
\begin{align*}
	&f(\widetilde  u_{jk},\sigma^2  \widetilde  M_{j,kk})\\
	&= \sqrt{\frac{2\sigma^2 \widetilde  M_{j, kk}}{\pi}} \exp\left(-\frac{\widetilde  u_{jk}^2}{2\sigma^2 \widetilde  M_{j,kk}}\right)
	+ \widetilde  u_{jk} \left(1-2\Phi\left(- \frac{\widetilde  u_{jk}}{\sqrt{\sigma^2 \widetilde  M_{j,kk}}} \right)\right),
\end{align*}
with $\Phi$ being the cumulative distribution function of a standard normal distribution. Here, $\det(B)$ and $\Tr(B)$ stands for the determinant and the trace of the matrix $B$. 

$\bullet$ When $q = 2$ and $m = 2$, it results in a multivariate normal density.
If $g$ is the multivariate normal density, we use $\mathcal{N}(0, \sigma^2I_r/\lambda_1 )$ instead, as the solution for $\sigma^2$ is simpler. One can consider we choose the tuning parameter to be $\lambda_1/\sigma^2$ instead of $\lambda_1$. 
Then, we obtain
\begin{align}
     \widehat  {\widetilde  u}_j  = 
    \widehat  {\widetilde M}_j
        \sum_{i=1}^n X_{ij} \widetilde  \omega_i',
\quad\text{and}\quad
\widehat  {\widetilde  M}_j = 
    \left(
        \sum_{i=1}^n 
        \left(
            \widetilde  \omega_i \widetilde  \omega_i' + \widetilde  V_w
        \right) + \lambda_1 I_r
    \right)^{-1}.\label{eqn:cavi-u-mvn}
\end{align}
\smallskip

{\bf 3. Updating $z$}

To solve $z$, we need to obtain $\widehat  { h} = (\widehat  h_1, \dots, \widehat  h_p)$, where for each $j\in \{1,\ldots,p\}$,
$\widehat  h_j = \log(\widehat  z_j / (1-\widehat  z_j))$. 
In Appendix \ref{derive-cavi}, we derive the solution for $\widehat  h_j$. If $g$ is the product of $r$ independent Laplace density, then
\begin{align}
\label{eqn:cavi-z}
\widehat  h_j & =  
\log \left(\frac{\alpha_1}{\alpha_2}\right) + r \log \left(\frac{\sqrt{\pi} \sigma \lambda_1}{\sqrt{2}}\right)
- \lambda_1 \sum_{k=1}^r f(\widetilde  u_{jk}, \sigma^2 \widetilde  M_{j,kk}) + \frac{\log \det(\widetilde  M_j)+1}{2} 
\nonumber \\
& \quad 
- \frac{1}{2\sigma^2} \sum_{i=1}^n 
\left[
 - 2 X_{ij} \widetilde  u_j \widetilde  \omega_i +\widetilde  u_j H_i \widetilde  u_j' + \Tr(\sigma^2  \widetilde  M_j H_i)
\right].
\end{align} 
If $g$ is the multivariate normal density, then
\begin{align}
\label{eqn:cavi-z-mvn}
  \widehat  h_j & = \log \left(\frac{\alpha_1}{\alpha_2} \right) + \frac{r\log\lambda_1}{2} - 
    \frac{\lambda_1}{2}
    \left(
        \widetilde  u_j \widetilde  u_j'  + \sigma^2 \Tr(\widetilde  M_j)
    \right) + \frac{\log \det (\widetilde  M_j) + 1}{2} \nonumber \\
    & \quad    - \frac{1}{2\sigma^2}
        \sum_{i=1}^n 
        \left(
        - 2X_{ij} \widetilde  u_j \widetilde  \omega_i' + \widetilde  u_j H_i \widetilde  u_j' + \Tr(\sigma^2 \widetilde  M_jH_i)
        \right).
\end{align}
\smallskip

{\bf 4. Updating $\widehat  \mu$ and $\widehat  M$}

As we obtained $\widehat  {\widetilde  u}$ and $\widehat  {\widetilde  M}$, then $\widehat  \mu$ and $\widehat  M$ can be solved accordingly.
Note that $\widetilde  w_i \sim \mathcal{N}(0, D)$. In the E-step, we also obtained $\widetilde  \omega_i$ and $V_\omega$. Thus, $D$ can be solved using $\widehat  D = \frac{1}{n} \sum_{i=1}^n \widetilde  \omega_i \widetilde  \omega_i' + \widetilde  V_\omega$, and $\widehat  \mu$ can be obtained by first solving $\widehat  u = \widehat  {\widetilde  u} \widehat  D_L$. Next, we apply the SVD to obtain $\widehat  A$. Last, we obtain $\mu$ using $\widehat  \mu = \widehat  u \widehat  A'$. 
$\widehat  M$ can be obtained similarly, i.e., $\widehat  M_j = \widehat  D_L \widetilde  M_j \widehat  D_L'$. 
\medskip

{\bf 5. Updating $\sigma^2$}

Recall that the prior $\sigma^2 \sim \text{InverseGamma}(\sigma_a, \sigma_b)$. If $g$ is the product of $r$ independent Laplace density, we obtain 
\begin{align}
\widehat \sigma^2 & = \argmin_{\sigma^2 \in (0, \infty)} 
\Bigg[
\sum_{j=1}^p z_j \Big\{
\frac{1}{2\sigma^2} \sum_{i=1}^n \Big( \widetilde  u_j H_i \widetilde  u_j' - 2X_{ij} \widetilde  u_j \widetilde  \omega_i \Big)
- \frac{r\log \sigma^2}{2} \nonumber\\
& \hspace{3cm}
+ \lambda_1 \sum_{k=1}^r f(\widetilde  u_{jk}, \sigma^2 \widetilde  M_{j,kk})
\Big\}
+\frac{(np+2\sigma_a+2)\log \sigma^2}{2} \label{eqn:cavi-sigma}\\
&\hspace{8cm}+ \frac{\Tr(X'X) + 2\sigma_b}{2\sigma^2} 
\Bigg].\nonumber
\end{align}
If $g$ is the multivariate normal density, we obtain
\begin{align}
\widehat  \sigma^2  = \frac{
\Tr(X'X) + \sum_{j=1}^p z_j \sum_{i=1}^n 
\left(
\widetilde  u_j H_i \widetilde  u_j'
- 2X_{ij} \widetilde  u_j \widetilde  \omega_i + \lambda_1 \widetilde  u_j \widetilde  u_j'
\right)  + 2\sigma_b
}{ 
np + 2(\sigma_a + 1)}.
\label{eqn:cavi-sigma-mvn}
\end{align}
Now, we summarize the PX-CAVI algorithm. 

\begin{algorithm}[ht]
\DontPrintSemicolon
   \vspace{0.1cm}
   
   \KwData{$X$, a $p \times n$ matrix, scaled and centered}
   \vspace{0.1cm}
   
  \KwInput{$\widehat  \mu^{(0)}$, $\widehat { M}^{(0)}$, $\widehat z^{(0)}$, $\widehat \sigma^{(0)}$ $r$, number of total iterations $T$, and the threshold $\delta$}
     \vspace{0.1cm}
     
{\bf For $t = 0, \dots, T-1$}:

  \begin{itemize}
  \item[--] Update $\widetilde  \omega^{(t+1)}$ and $\widetilde  V_w^{(t+1)}$ using (\ref{eqn:cavi-w})
  \item[--] If $g$ is the product of $r$ independent Laplace density
  \begin{itemize}
  	\item[--] Update $ \widetilde  u^{(t+1)}$ and $\widetilde {M}^{(t+1)}$ using (\ref{eqn:cavi-u}) and (\ref{eqn:cavi-Xi}) 
	  \item[--] Update $ h^{(t+1)}$ using (\ref{eqn:cavi-z}) and then obtain 
  $\widehat  z^{(t+1)}$
  \item[--] Update $\sigma^{(t+1)}$ from (\ref{eqn:cavi-sigma}) 
\end{itemize}
\item[--] If $g$ is the multivariate normal density
  \begin{itemize}
  \item[--] Update $\widetilde  u^{(t+1)}$ and $\widetilde {M}^{(t+1)}$ using (\ref{eqn:cavi-u-mvn})
  \item[--] Update $ h^{(t+1)}$ using (\ref{eqn:cavi-z-mvn}) and then obtain 
  $\widehat  z^{(t+1)}$
  \item[--] Update $\sigma^{(t+1)}$ using (\ref{eqn:cavi-sigma-mvn})
  \end{itemize}
    \item[--] Obtain $D^{(t+1)}$, $u^{(t+1)}$, and $\widehat  M^{(t+1)}$
  \item[--] Apply SVD to obtain $A^{(t+1)}$ and then obtain $\mu^{(t+1)}$
\end{itemize}
   \vspace{0.1cm}
   
{\bf Stop}: If $\max\left(\big\|\mu^{(t+1)} \mu^{{(t+1)}'} - \mu^{(t)} \mu^{{(t)}'}\big\|_F, \| z^{(t+1)} -z^{(t)}\|_1\right) \leq \delta$
   
 \KwOutput{$\widehat  P(\theta)$. }
    \vspace{0.1cm}
    
\caption{The PX-CAVI algorithm}
\label{px-cavi}
\end{algorithm}


\section{Asymptotic properties}
\label{sec:postCR}

This section studies the asymptotical properties of the posterior in (\ref{posterior}) and the variational posterior in (\ref{eqn:MF-posterior}).
We work with the subset selection prior, which includes the spike and slab prior in (\ref{spike-and-slab-prior}) as a special case, which is constructed as follows: First, a number $s$ is chosen from a prior $\pi$ on the set $\{0, \dots, p\}$. Next, a set $S$ is chosen uniformly from the set $\{1, \dots, p\}$ such that its cardinality $|S| = s$. 
Last, conditional on $S$, if $j \in S$, then the prior for $\theta_j$ is chosen to be $\int_{A \in V_{r,r}} g(\theta_j|\lambda_1, A) d\Pi(A)$; if $j \not\in S$, then $\theta_j$ is set to $0_r'$.
The prior is given as follows:
\begin{align}
\label{spike-and-slab-v2}
    \pi(\theta, S|\lambda_1)
    \propto 
    \pi(|S|)
    \frac{1}{{p \choose |S|}}
    \prod_{j \in S}
     \int_{A \in V_{r,r}} g(\theta_{j}| \lambda_1, A)\pi(A) dA
    \prod_{j \not\in S} 
    \delta_0(\theta_j).
\end{align}
Note that (\ref{spike-and-slab-prior}) is a special case of (\ref{spike-and-slab-v2}) when $\pi(|S|)$ is the beta-binomial distribution. That is, $s|\kappa \sim \text{binomial}(p, \kappa)$ and $\kappa \sim \text{Beta}(\alpha_1, \alpha_2)$. 

In the next subsection, we will study the theoretical properties of the posterior with the subset selection prior. 
Before we proceed, some notations need to be introduced.
Let $\lesssim$ (resp. $\gtrsim$) stand for inequalities up (resp. down) to a constant, $a \asymp b$ stand for $C_1a \leq b \leq C_2a$ with positive constants $C_1 < C_2$, and
$a \ll b$ stand for $a/b \to 0$.
We denote $\|b\|_2$ as the $\ell_2$-norm of a vector $b$ and $\|B\|$ as the spectrum norm of a matrix $B$.
The true value of an unknown parameter $\vartheta$ is denoted by $\vartheta^\star$.

\subsection{Contraction rate of the posterior}

We study the dimensionality and the contraction rate of the posterior distribution. In this study, we assume $r$ is unknown and $\sigma^2$ is fixed. Three assumptions are needed to obtain the rate.

\begin{assumption}[Priors for $s$ and $r$] 
\label{assumps}
For positive constants $a_1$, $a_2$, $a_3$, and $a_4$, assume
$$p^{-a_1} \lesssim \pi(s)/\pi(s-1) \lesssim p^{-a_2}\quad \text{and}\quad \exp(-a_3 r) \lesssim \pi(r) \lesssim \exp(- a_4 r).$$
\end{assumption}

The above assumption impose conditions on the tails of the priors $\pi(s)$ and $\pi(r)$.
The first condition also appears in the study of the sparse linear regression model \citep[e.g.][]{cast15, martin17, ning20}. It assumes that the logarithm of the ratio between $\pi(s+h)$ and $\pi(s)$ is in the same magnitude as $-h\log p$.
When $h$ increases, the assigned probability on $s+h$ decays exponentially fast. 
The beta-binomial prior mentioned above satisfies this condition if one chooses, for example, $\alpha_1 = 1$ and $\alpha_2 = p^{\nu} + 1$ for any $\nu > \log \log p/ \log p$.
The second condition is similar to that in \citet{pati14}. It assumes the tail of $\pi(r)$ should decay exponentially fast; the Poisson distribution satisfies this condition.

\begin{assumption}[Bounds for $\lambda_1$]
\label{assump:lambda1}
For positive constants $b_1, b_2,$ and $b_3$, assume
    $$
    b_1 \sqrt{\frac{n}{p^{b_2/r^\star}}} 
    \leq \lambda_1 \leq 
    b_3 \sqrt{n\log p}.
    $$
\end{assumption}

Assumption \ref{assump:lambda1} provides the permissible region for $\lambda_1$.
If $\lambda_1$ is too large, it introduces an extra shrinkage effect on large signals; if it is too small, the posterior will contract at a slower rate.
Our upper bound is of the same order as that in \citet{cast15}, where they studied the sparse linear regression model. But the lower bounds are different. Ours is bigger; it can go to 0 very slowly if $r^\star$ is close to $\log p/\log n$. 

\begin{assumption}[Bounds for $r^\star$ and $\theta^\star$]
\label{assump:truevalue}
For some positive constant $b_2$, $b_4$, and $b_5$, $r^\star \leq b_2\log p/\log n$,
    $\|\theta^\star\| \geq b_4$ and $\|\theta^\star\|_{1,1} \leq b_5 s^\star \log p/\lambda_1$ if $m = 1$ and $1\leq q\leq 2$ and $\|\theta^\star\|^2 \leq b_5 s^\star \log p/\lambda_1$ if $m = 2$ and $q = 2$.
\end{assumption}

Assumption \ref{assump:truevalue} requires the true values of $\theta$ and $r$ being bounded. $r^\star$ cannot be too large. If $\log p/\log n \lesssim r^\star \lesssim \log p$, then the rate obtained in Theorem \ref{Thm: Post-cont-rate} will be slower, i.e., $O(\sqrt{r^\star s^\star \log p/n}$). The bounds for $\|\theta^\star\|$ essentially control the largest eigenvalue, as $\|\theta^\star\|^2 + \sigma^2$ is the largest eigenvalue of $\Sigma^\star$. It cannot be either too big or too small. 

We now present the main theorem.
\begin{theorem}
\label{Thm: Post-cont-rate}
For the model in (\ref{model}) and the subset selection prior in (\ref{spike-and-slab-v2}), if Assumptions \ref{assumps}-\ref{assump:truevalue} hold, then for sufficiently large constants $M_1$, $M_2$, and $M_3 \geq M_2/b_4$, as $n$ goes to infinity,
\begin{eqnarray}
    & \mathbb{E}_{f^\star} \Pi(\theta: |S| > M_1 s^\star| X) \to 0, 
	\label{Thm-1:eqn-3}& \\
	& \mathbb{E}_{f^\star} \Pi(\|\Sigma - \Sigma^\star\| \geq M_2\epsilon_n| X) \to 0,
	\label{Thm-1:eqn-1}& \\
	&\mathbb{E}_{f^\star}  
	\Pi \left(\left\|UU' - U^\star {U^\star}'\right\| \geq M_3\epsilon_n| X \right) \to 0,
	\label{Thm-1:eqn-2}&
\end{eqnarray}
where $\epsilon_n = \sqrt{s^\star \log p/n}$.
\end{theorem}

In Theorem \ref{Thm: Post-cont-rate}, we derive the posterior contraction rate under the spectrum loss.
The minimax rates for using the spectrum loss have been studied by \citet{cai15}.
Consider the parameter space 
$$
\Theta_0(s, p, r, \overline\rho) = \Big\{
\Sigma: 0 \leq \|\theta_{\cdot r}\|_2^2 \leq \|\theta_{\cdot 1}\|_2^2 \leq \overline\rho,\; U \in V_{r, r},\; |S| \leq s 
\Big\},
$$
the minimax rate of estimating $\Sigma$ for $r \leq s \leq p$ is 
$
\sqrt{\frac{(\overline\rho + 1)s}{n} \log \left( \frac{ep}{s} \right) + \frac{\overline\rho^2 r}{n}} \wedge \overline\rho
$.
Comparing it to the rate we obtained, assuming $\overline\rho$ is fixed and $s \geq r$, our rate is suboptimal as the log factor in our rate is $\log p$ but in the minimax rate, it is $\log (p/s)$. \citet{cai15} also provided the minimax rate for the projection matrix. Assuming a more restrictive parameter space $\Theta_1(s, p, r, \overline{\rho}, \tau)$,
$$
\Theta_1(s, p, r, \overline\rho, \tau) = \Big\{
\Sigma: \overline\rho/\tau \leq \|\theta_{\cdot r}\|_2^2 \leq \|\theta_{\cdot 1}\|_2^2 \leq \overline\rho,\; U \in V_{r, r},\; |S| \leq s
\Big\},
$$
the minimax rate is $\sqrt{\frac{(\overline\rho + 1)s}{n\overline\rho^2} \log \left(\frac{ep}{s}\right)} \wedge 1$. Again, if $\overline\rho$ is fixed, the rate we obtained is suboptimal.

One may ask if we could obtain the same rate as that in Theorem \ref{Thm: Post-cont-rate} if we use the Frobenius norm as the loss function (in short, Frobenius loss). This is in fact possible, and the proof can simply follow the argument in \citet{gao15}. However, one needs to impose a lower bound for $\|\theta{\cdot r}\|_2^2$. Although in practice, the lower bound can be introduced through the prior, e.g., using a truncated prior, the exact value is hard to determine. Thus, we did not choose this prior.


\subsection{Contraction rate of the variational posterior}

We study the contraction rate of the variational posterior in (\ref{eqn:MF-posterior}). Recent studies on this topic have provided exciting results of the variational method and developed useful tools for studying their theoretical properties \citep[e.g.][]{ray20, wang19, yang20, zhang20}. 
\citet{ray20} and \citet{yang20} studied the spike and slab posterior with the linear regression model and obtained a (near-)optimal rate for their posterior. \citet{zhang20} proposed a general framework for deriving the contraction rate of a variational posterior. We derive the rate by directly applying this general framework, as our variational posterior is intractable, and using a direct argument (e.g., those in the linear regression model) is impossible. Theorem \ref{Thm: variational-post-contr-rate} shows that the rate of the variational posterior is also $\epsilon_n$ (but with a larger constant). Proofs of the theorem are provided in the supplemental material. \\

\begin{theorem}
\label{Thm: variational-post-contr-rate}
With the model (\ref{model}) and the subset selection prior (\ref{spike-and-slab-v2}), if $\widehat  P(\theta) \in \mathcal{P}^{\MF}$ and Assumptions \ref{assumps}-\ref{assump:truevalue} hold, then for large constants $M_4$ and $M_5$, as $n$ goes to infinity,
\begin{align}
    & \widehat {Q}(\|\Sigma - \Sigma^\star\| \geq M_4\epsilon_n|X) \to 0,
    \label{thm5.1-1}\\
    & \widehat {Q}(\|UU' - U^\star{U^\star}'\| \geq M_5 \epsilon_n|X) \to 0.
    \label{thm5.1-2}
\end{align}
\end{theorem}


\section{The PX-EM algorithm}
\label{sec:EM}

The EM algorithm is another popular algorithm that is used in Bayesian high-dimensional analysis.
In this section, to understand the strength of PX-CAVI, we also develop its EM analog, referred to as the PX-EM algorithm. The parameter expansion steps for the PX-EM algorithm mirror those used in the PX-CAVI algorithm.
The PX-EM algorithm requires us to use the continuous spike and slab prior, which is 
\begin{align}
\pi(\theta, S|\lambda_1, \lambda_0) \propto \int_A \prod_{j=1}^p \Big[
\gamma_j g(\theta_j| \lambda_1, A, r) + (1-\gamma_j) g(\theta_j|\lambda_0, A, r)
\Big] \pi(A) dA,
\end{align}
where $\lambda_0 \gg \lambda_1$. 
By comparing to (\ref{spike-and-slab-prior}), the Dirac measure is replaced by the continuous density with a large variance. 
The priors for the rest parameters remain the same.

Our PX-EM algorithm contains two steps: E-step and M-step. In the E-step, expectations are taken with respect to both $w$ and $\bgamma$. We then obtain 
\begin{eqnarray}
\label{eqn:EM-w}
    & w_i|\theta^{(t)}, X \sim \mathcal{N}(\omega_i, V_w), \\
    &\gamma_j \sim {\text{Bernoulli}}(\widetilde \gamma_j),
     \label{eqn:EM-gamma}
\end{eqnarray}
where
$\theta^{(t)}$ and $\kappa^{(t)}$ are the estimated values of $\theta$ and $\kappa$ from the $t$-th iteration
and 
\begin{eqnarray}
\label{eqn:EM-omega-v}
& V_w = \sigma^2 ({\theta^{(t)}}'\theta^{(t)} + \sigma^2 I_r)^{-1}, \quad
\omega_i = \sigma^{-2} V_w {\theta^{(t)}}' X_{i},& \\
\label{eqn:EM-gamma-tilde}
& \widetilde \gamma_j^{(t)} = P(\gamma_j=1|\theta^{(t)}, \kappa^{(t)}, X) = \frac{a_j^{(t)}}{a_j^{(t)}+b_j^{(t)}},& 
\end{eqnarray}
where $a_j^{(t)} = \exp( - \lambda_1 \|\theta^{(t)}_j\|_q^m + \log \kappa^{(t)})$
and 
$b_j^{(t)} = \exp( - \lambda_0 \|\theta^{(t)}_j\|_q^m + \log (1-\kappa^{(t)}))$.

To obtain the objective function, we first apply parameter expansion to the likelihood, same as that in the PX-CAVI algorithm. The expanded parameter becomes $\widetilde  \beta = \beta D = \theta A D$. The spike and slab prior is then directly applied on $\widetilde  \beta$. The objective function is given by
$
Q(\widetilde  \beta, \kappa| \theta^{(t)}, A^{(t)}, D^{(t)}, \kappa^{(t)})
$, where
\begin{equation}
\begin{split}
\label{obj-fn}
Q  & = \mathbb{E}_{w, \bgamma|\theta^{(t)}, \kappa^{(t)}} \log \pi(\widetilde  \beta, w|X)
\\
& = C - \sum_{j=1}^p \left(
        \frac{1}{2\sigma^2} \left\|M_L \widetilde  \beta_j' - d_j\right\|_2^2 + 
        (\widetilde \gamma_j \lambda_1 + (1-\widetilde \gamma_j) \lambda_0) \|\widetilde \beta_j\|^m_q
    \right)     \\
    & \quad + 
    \left(\|\widetilde \bgamma\|_1 + \alpha_1 - 1\right)\log \kappa
    + \left(p - \|\widetilde \bgamma\|_1 + \alpha_2 - 1\right)\log (1-\kappa),
    \end{split}
\end{equation}
where $C$ is a constant,
$M_L$ is the lower triangular part from the Cholesky decomposition, $M = \sum_{i=1}^n \omega_i \omega_i' + nV_w$, and $d_j = {M_L}^{-1} \sum_{i=1}^n \omega_i X_{ij}$.

In the M-step, we maximize the objective function and obtain 
\begin{align}
\label{eqn:EM-beta}
    \widehat {\widetilde \beta}_j 
    & = \argmin_{\widetilde \beta_j}
    \left\{
    \frac{1}{2\sigma^2} 
    \left\|M_L \widetilde \beta_j' - d_j \right \|_2^2 + 
    \text{pen}_j \|\widetilde \beta_j\|_q^m
    \right\},\\
\label{eqn:EM-kappa}
    \widehat \kappa & = \frac{\alpha_1 + \|\widetilde \bgamma\|_1 - 1}{p + \alpha_1 + \alpha_2 - 2},
\end{align}
where $\text{pen}_j = \widetilde  \gamma_j \lambda_1 + (1-\widetilde \gamma_j) \lambda_0$.
Then $\widehat \theta$ is obtained using $\widetilde  \beta = \theta A D_L$, 
where $\widehat  D = \frac{1}{n} \sum_{i=1}^n{\omega_i \omega_i'} + V_w$
and $\widehat  A$ is obtained by applying the SVD on the matrix $\widehat  {\widetilde  \beta} \widehat  D_L^{-1}$.

In (\ref{eqn:EM-beta}), we choose $m = 1$ and let $q = 1$ and $2$. When $q = 1$, the expression is similar to that of the adaptive lasso \citep{zou06}. When $q = 2$, the penalty term is then similar to it in the group lasso method \citep{ming06}. 
Despite those similarities, the tuning parameter in (\ref{eqn:EM-beta}) can be updated during each EM iteration; however, in both of the two aforementioned literature, their tuning parameters are chosen to be fixed values. The benefit of allowing the tuning parameter to update is explored by 
\citet{rockova18b}, which studied the sparse normal mean model. 

Last, we obtain
\begin{align}
\label{eqn:EM-sigma}
\widehat  \sigma^2 = \frac{
\Tr(X'X) - 2 \sum_{j=1}^p d_j M_L \theta_j' + \sum_{j=1}^p \theta_j M \theta_j' + 2\sigma_b
}{
np + 2(\sigma_a + 1).
}
\end{align}

\begin{algorithm}[ht]
\DontPrintSemicolon
   \KwData{$X$, a $p \times n$ matrix, centered and scaled}
      \vspace{0.1cm}
   
  \KwInput{$\theta^{(0)}$, $\sigma^{(0)}$, $r$, number of total iterations $T$, and the threshold $\delta$}
     \vspace{0.1cm}
  
{\bf For $t = 0, \dots, T-1$, repeat}:

  \begin{itemize}
  \item[--] Update $\omega^{(t+1)}$ and $V_w^{(t+1)}$ from (\ref{eqn:EM-w}) and ${\widetilde \bgamma}^{(t+1)}$ from (\ref{eqn:EM-gamma});
  \item[--] Update ${\widetilde \beta}_j^{(t+1)}$ from (\ref{eqn:EM-beta})
  \item[--] Update $\kappa^{(t+1)}$ from (\ref{eqn:EM-kappa})
  \item[--] Update $D^{(t+1)}$ and $A^{(t+1)}$ and then obtain $\theta^{(t+1)}$
  and $U^{(t+1)}$
  \item[--] Update $\sigma^{(t+1)}$ from (\ref{eqn:EM-sigma})
  \item[--] Evaluate the objective function ${Q}^{(t+1)}$ in (\ref{obj-fn})
\end{itemize}

{\bf Stop}: if  $\big|\log {Q}^{(t+1)} - \log {Q}^{(t)}\big|\leq \delta$
   \vspace{0.1cm}

 \KwOutput{$\widehat \theta = \theta^{(t+1)}$, $U = \widehat  U^{(t+1)}$, 
 $\widehat {\bgamma} =\widehat {\widetilde \bgamma}^{(t+1)}$, and 
 $\widehat  \sigma = \widehat  \sigma^{(t+1)}$.}
   \vspace{0.1cm}

\caption{The PX-EM algorithm}
\label{px-em}
\end{algorithm} 

We conclude this section by offering theoretical justification for utilizing parameter expansion to accelerate the convergence speed of the EM algorithm. We observed that the convergence speed improves with both parameter expansions.
Intuitively, by \citet{dempster77}, the speed of convergence is determined by the largest eigenvalue of $S(\Delta) = I^{-1}_{com}(\Delta) I_{obs}(\Delta)$, where
\begin{align}
\label{info-mtx}
I_{obs}(\Delta) = -\frac{\partial^2 \log(\Delta|X)}{\partial \Delta \partial \Delta'}\Bigg|_{\Delta = \Delta^\star}
\quad\text{and}\quad
I_{com}(\Delta) = -\frac{\partial^2 Q(\Delta|\Delta)}{\partial \Delta \partial \Delta'}\Bigg|_{\Delta = \Delta^\star}.
\end{align}
We denote $\Delta$ as the collection of all the unknown parameters and $\Delta^\star$ as the true values.
Let $\Psi$ be the expanded parameter and $\widetilde  \Delta = (\Delta, \Psi)$.  We found that the largest eigenvalue of $S(\widetilde  \Delta)$ is bigger than that of $S(\Delta)$. Thus, the convergence speed is increased. In Lemma \ref{lemma-convergence-EM}, we provide a formal statement of this result. Proof of Lemma \ref{lemma-convergence-EM} is provided in the supplementary material. 
\begin{lemma}
\label{lemma-convergence-EM}
Given that the PX-EM algorithm converges to the posterior mode, both parameter expansions speed up the convergence of the original EM algorithm.
\end{lemma}

\section{Simulation study}
\label{sec:simstudy}

In this section, we conduct four simulation studies to evaluate the performance of our proposed PX-CAVI algorithm. Firstly, we compare the use of a product of Laplace density (i.e., $q = 1$ and $m=1$ in $g$ (\ref{prior-beta})) with the multivariate normal density (i.e., $q = 2$ and $m = 2$ in $g$ (\ref{prior-beta})) within the PX-CAVI algorithm. Next, we compare the PX-CAVI algorithm with the PX-EM algorithm. Additionally, we introduce the batch PX-CAVI algorithm, which does not require $\theta$ to be jointly row-sparse, and compare it with two other penalty methods for SPCA and the conventional PCA.
In the final study, we assume that $r$ is unknown and demonstrate that the algorithm is less sensitive to the choice of $r$.
Throughout all the studies, we set $\sigma^2$ to be fixed. However, in the $\mathsf{R}$ package we provided, it has the capability to estimate $\sigma^2$ automatically.

The dataset is generated as follows:
First, given $r^\star$, $s^\star$, and $p$, we generate $U^\star$ using the $\mathsf{randortho}$ function in $\mathsf{R}$.
Next, we set $\sigma^2 = 0.1$ and choose the diagonal values of $\Lambda^\star$ to be an equally spaced sequence from 10 to 20 (i.e., the largest value is 20 and the smallest value is 10); however, in the first study, we will choose different values for $\Lambda^\star$; see Section \ref{sec:sim-1} for details.
Last, we obtain $\Sigma^\star = U^\star \Lambda^\star {U^\star}' + \sigma^2 I_p$ and generate $n = 200$ independent samples from $\mathcal{N}(0, \Sigma^\star)$. Then, the dataset is an $n \times p$ matrix. 
For each simulated dataset, we obtain the following quantities: the Frobenius loss of the projection matrix $\|\widehat  U\widehat  U' - U^\star {U^\star}'\|_F$, 
the percentage of misclassification also known as the average Hamming distance
$\|\widehat  z - \bgamma^\star\|_1/p$,  the false discovery rate (FDR), and the false negative rate (FNR). 
The hyperparameters in the prior are chosen as follows: $\lambda_1 = 1$, $\alpha_1 = 1$, $\alpha_2 = p + 1$, $\sigma_a = 1$, and $\sigma_b = 2$. Also, we set the total iterations $T = 100$, $\iota = 0.1$, and the threshold $\delta = 10^{-4}$. To determine whether $\gamma_j = 1$ or $0$, we choose the threshold to be $0.5$.

\subsection{On choosing the initial values for PX-CAVI and PX-EM}
\label{sec:initial-values}

Before presenting the simulation results, it is necessary to discuss how we obtained the initial values for the PX-CAVI algorithm which is the same for the batch PX-CAVI algorithm, and the initial values for the PX-EM algorithm. 
We carefully explored different choices of initial values and found that the PX-CAVI algorithm exhibits robustness against variations in the initial values. Consequently, the algorithm is not overly sensitive to the specific choices of initializations.Therefore, we estimated $\widehat \mu^{(0)}$ using the conventional PCA and set $\widehat  z^{(0)} = \mathbbm{1}_p'$. For $\widehat M_j^{(0)}$, we let it be an identity matrix times a small value (i.e., $10^{-3}$). Finally, for $(\widehat \sigma^{(0)})^2$, we chose it to be the smallest eigenvalue of the Gramian matrix $X'X/(n-1)$. 

The PX-EM algorithm is more sensitive to poor initializations than the PX-CAVI algorithm. To address this concern, we employed two strategies aimed at alleviating this issue. The first one is proposed by \citet{rockova14}, which we replaced (\ref{eqn:EM-gamma}) with its tempered version given by 
\begin{align}
\label{eqn:termpered-gamma}
\widetilde \gamma_j^{(t)} = P(\gamma_j=1|\theta^{(t)}, X) = \frac{\left(a_j^{(t)}\right)^\iota}{\left(a_j^{(t)}\right)^\iota
+ \left(b_j^{(t)}\right)^\iota}.
\end{align}
where $\iota < 1$ is fixed. 
In the simulation study, we fix $\iota = 0.1$.
Another strategy we adopted is the path-following strategy proposed by \citet{rockova16}. First, we chose a vector containing a sequence of values of $\lambda_0$, $\{\lambda_0^{(1)}, \dots, \lambda_0^{(I)}$\}, where $\lambda_0^{(1)} = \lambda_1 + 2\sqrt{\rho_{\min}}$ with $\rho_{\min}$ being the smallest eigenvalue of $X'X/(n-1)$, and $\lambda_0^{(I)} = p^2\log p$. Next, we obtained an initial value of $\theta$ using the conventional PCA and repeated the following process: At $i$-th step, set $\lambda_0 = \lambda_0^{(i)}$ and chose the input values as their output values obtained from the $(i-1)$-th step. We repeated this $I$ times until all the values in that sequence of $\lambda_0$ are used. Finally, the values output from the last step are used as the initial values for the PX-EM algorithm. As can be seen, comparing to the PX-CAVI algorithm, obtaining the initial values of the PX-EM algorithm takes a much longer time. 

\subsection{Laplace density vs normal density}
\label{sec:sim-1}

Let $r^\star =1$, then $g$ is the Laplace distribution ($m =1, q = 1$) and the normal distribution ($m = 2, q = 2$). We conducted  simulation studies of the PX-CAVI algorithm and compared the use of two distributions. We chose $\|\theta^\star\|^2 \in \{1, 3, 5, 10, 20\}$ and $p \in \{100, 1000\}$. For each setting,1000 datasets are generated. 
Simulation results are provided in Table \ref{sim-1}.

\begin{table}[h!]
\begin{adjustwidth}{-.1in}{-.1in}  
\centering
\caption{\small Simulation results of the PX-CAVI algorithm using the Laplace and Normal densitis. We fixed $n = 200$, $s^\star = 20$ and $r^\star = 1$ and chose $p \in \{100, 1000\}$ and $\|\theta^\star\|^2 \in \{1, 3, 5, 10, 20\}$. For each setting, we ran 1000 simulations and computed the average values of the Frobenius loss of the projection matrix, the percentage of misclassification, FDR, and FNR.}
\label{sim-1}
{ \small
\begin{tabular}{cc|cccc||cccc}
\Xhline{2\arrayrulewidth}
 &  & \multicolumn{4}{c||}{$p = 100$} & \multicolumn{4}{c}{$p = 1000$} 
\\\Xhline{2\arrayrulewidth}
$\|\theta^\star\|^2$ & Prior & Frob & Misc(\%) & FDR & FNR & Frob & Misc(\%) & FDR & FNR \\ \hline
\multirow{2}{*}{$1$} & Normal & 0.156 & 2.4 & 0.000 & 0.026 & 0.174 & 0.3 & 0.000 & 0.003\\
                           & Laplace & 0.156 & 2.4 & 0.000 & 0.026 & 0.190 & 0.3 & 0.000 & 0.003\\ \hline
\multirow{2}{*}{$3$} & Normal & 0.076 & 1.4 & 0.000 & 0.015 & 0.082 & 0.1 & 0.000 & 0.001\\
                           & Laplace & 0.076 & 1.4 & 0.000 & 0.015 & 0.088 & 0.2 & 0.000 & 0.002\\ \hline
\multirow{2}{*}{$5$} & Normal & 0.055 & 1.0 & 0.000 & 0.012 & 0.059 & 0.1 & 0.000 & 0.001\\
                           & Laplace & 0.055 & 1.0 & 0.000 & 0.012 & 0.062 & 0.1 & 0.000 & 0.001\\ \hline
\multirow{2}{*}{$10$} & Normal & 0.036 & 0.8 & 0.000 & 0.008 & 0.038 & 0.1 & 0.000 & 0.001\\
                             & Laplace & 0.036 & 0.8 & 0.000 & 0.008 & 0.039 & 0.1 & 0.000 & 0.001\\ \hline
\multirow{2}{*}{$20$} & Normal & 0.024 & 0.5 & 0.000 & 0.006 & 0.026 & 0.1 & 0.000 & 0.001\\
                            & Laplace & 0.024 & 0.5 & 0.000 & 0.006 & 0.026 & 0.1 & 0.000 & 0.001\\ \hline
\Xhline{2\arrayrulewidth}    
\end{tabular}  
}
\end{adjustwidth}
\end{table}

From Table \ref{sim-1}, we observed the following results:
For $p = 100$, there is no significant difference between using the normal and the Laplace densities, as their results are similar. However, when $p = 1000$, using the normal density yields better results, as indicated by the smaller average value of the Frobenius loss of the projection matrix.
In the case of $p = 1000$, the normal density outperforms the Laplace density in estimating weaker signals (e.g., observed in the Frobenius loss when $|\theta^\star| = 1$). The computational speed using the normal density is faster than the Laplace density. This is because when choosing the Laplace density, the algorithm needs to solve the two nonlinear functions (\ref{eqn:cavi-u}) and (\ref{eqn:cavi-Xi}) in each iteration. The computational speed notably increases, particularly when $r > 2$ using the Laplace density, and solving the two equations (\ref{eqn:cavi-u}) and (\ref{eqn:cavi-Xi}) becomes more challenging. Based on these findings, we recommend using the multivariate normal density, especially when the rank $r$ is large, as it provides improved performance and computational efficiency in comparison to the Laplace density.

\subsection{Comparison between PX-CAVI and PX-EM}

In this study, we compare the PX-CAVI algorithm with the PX-EM algorithm. Two options for $q$ in (\ref{eqn:EM-beta}) are considered for the PX-EM algorithm: $q = 1$ representing the $\ell_1$-norm, and $q = 2$ representing the $\ell_2$-norm. We observed that the algorithm using the $\ell_1$-norm outperforms the one using the $\ell_2$-norm in terms of parameter estimation  (see the simulation result in the Supplementary Material). Henceforth, we utilized the $\ell_1$-norm. The true parameter values were chosen as follows: We fixed $s^\star = 20$, and $r^\star = 2$ and chose $q = 1$, $s^\star \in \{10, 40, 70, 150\}$, $r^\star \in \{1,3,5\}$, and $p \in \{500, 1000, 2000, 4000\}$.
We ran both the PX-CAVI and the PX-EM algorithms. The results are given in Table \ref{sim-2}. As we mentioned in Section \ref{sec:initial-values}, choosing the initial values for the PX-EM algorithm takes a longer time, and thus, we were only able to run 100 simulations. For the PX-CAVI, the result is based on 1000 simulations. 

We remark two findings in Table \ref{sim-2}. First, in general, the PX-CAVI algorithm is better than the PX-EM algorithm in both parameter estimation and variable selection. When $s^\star$ and $r^\star$ are large, the PX-CAVI algorithm is more accurate. Although it seems that when $s^\star$ and $r^\star$ are small (e.g., $s^\star = 10$ and $r^\star = 1$ and $s^\star = 40$ and $r^\star = 1$), the Frobenius loss and the percentage of misclassification are bigger in the PX-CAVI algorithm than the PX-EM algorithm. However, the standard errors associate with the Frobenius loss when $s^\star = 10$ and $r^\star = 1$ is 0.011 and $s^\star = 40$ and $r^\star = 1$ is 0.015. For the percentage of misclassification, the standard errors are 0.1 when $s^\star = 10$ and $r^\star = 1$ and 0.2 when $s^\star = 40$ and $r^\star = 1$. Consequently, the observed differences between the two algorithms are insignificant. Our second notable finding is that both algorithms effectively control the FDR. However, the PX-CAVI algorithm exhibits better control over the FNR, resulting in more accurate and desirable variable selection outcomes.
\begin{table}[h!]
\begin{adjustwidth}{-.9in}{-.9in}  
\centering
\caption{\small Simulation results of the PX-CAVI and the PX-EM algorithms. We fixed $n = 200$ and chose $s^\star \in \{10, 20, 40, 70, 150\}$, $r^\star \in \{1, 2, 3, 5\}$, and $p \in \{500, 1000, 2000, 4000\}$. For each setting, we ran $100$ simulations for the PX-EM and $1000$ simulations for the PX-CAVI. We computed the average values of the Frobenius loss of the projection matrix, the percentage of misclassification, FDR, and FNR.}
\label{sim-2}
{ 
\small
\begin{tabular}{ccc|cc|cc|cc|cc}
\Xhline{2\arrayrulewidth}
\multicolumn{3}{c}{} & \multicolumn{2}{c}{\bf Frobenius loss} & \multicolumn{2}{c}{\bf Misc (\%)}
& \multicolumn{2}{c}{\bf FDR} & \multicolumn{2}{c}{\bf FNR}
\\\Xhline{2\arrayrulewidth}
    $p$ & $s^\star$ & $r^\star$ &  PX-CAVI & PX-EM
    &  PX-CAVI & PX-EM &  PX-CAVI & PX-EM &  PX-CAVI & PX-EM\\
\hline
    $1000$ & $10$ & $1$  &
    0.025 & 0.024 &
    0.1 &  0.1 &
    0.000 & 0.001 & 0.001 & 0.001\\
    $1000$ & $10$ & $3$ &
    0.039 & 0.040  &
    0.1 &  0.0 &
    0.000 & 0.000 & 0.000 & 0.000\\
    $1000$ & $10$ & $5$ &
    0.043 & 0.043 &
    0.1 & 0.0 
    & 0.000 & 0.000 & 0.000 & 0.000\\ 
    \hline
    $1000$ & $40$ & $1$  &
    0.061  & 0.054  &
    0.5 & 0.4  &
    0.000 & 0.001 & 0.006 & 0.004\\
    $1000$ & $40$ & $3$ &
    0.092 & 0.128 &
    0.0 &  0.1
    & 0.000 & 0.000 & 0.000 & 0.001\\
    $1000$ & $40$ & $5$ &
    0.113 & 0.128 &
    0.0 &  0.1 & 
    0.000 & 0.000 & 0.000 & 0.001\\
    \hline
    $1000$ & $70$ & $1$  &
    0.089 & 0.093  &
    1.2 & 1.2 &
    0.000 & 0.000 & 0.016 & 0.013\\
    $1000$ & $70$ & $3$ &
    0.126 & 0.214  &
    0.1 &  0.4  &
    0.000 & 0.000 & 0.001 & 0.005\\
    $1000$ & $70$ & $5$ &
    0.155 & 0.212 &
    0.0 &  0.1 
    &0.000 & 0.000 & 0.000 & 0.002\\   
    \hline
    $1000$ & $150$ & $1$  &
    0.145 & 0.155 &
    3.5 &  3.7
    & 0.000  & 0.000 & 0.052 & 0.042\\
    $1000$ & $150$ & $3$ &
    0.194 & 0.463 &
    0.3 &  2.6 
    & 0.000 & 0.000 & 0.010 & 0.029\\
    $1000$ & $150$ & $5$ &
    0.231 & 0.520  &
    0.0 &  1.5 &
    0.000 & 0.000 & 0.002 & 0.017\\   
\hline 
    $500$ & $20$ & $2$  &
    0.054 & 0.067 &
    0.0 &  0.1 &
    0.000 & 0.000 & 0.001 & 0.001\\
    $1000$ & $20$ & $2$ &
    0.054 & 0.072 &
    0.0 &  0.1 &
    0.000 & 0.000 & 0.000 & 0.001\\
    $2000$ & $20$ & $2$ &
    0.055 & 0.063 &
    0.0 &  0.0  & 
    0.000 & 0.000 & 0.000 & 0.001\\
    $4000$ & $20$ & $2$  &
    0.055 & 0.057 &
    0.0 & 0.0 & 
    0.000 & 0.000 & 0.000 & 0.001\\
\Xhline{2\arrayrulewidth}    
\end{tabular}  
}
\end{adjustwidth}
\end{table}

\subsection{The batch PX-CAVI vs other SPCA algorithms}
\label{sim-3}
The PX-CAVI algorithm assumes $\theta$ to be jointly row-sparse. In the Supplementary Material, we propose the batch PX-CAVI algorithm, which relaxes this assumption, allowing each principal component to have identical support. The batch PX-CAVI algorithm updates the coordinates belonging to the same row simultaneously. To evaluate the performance of the batch PX-CAVI algorithm, we compare it with two other popular algorithms for SPCA: the elastic net method proposed by \citet{zou06} and the robust SPCA method proposed by \citet{erichson20}. Both of these methods are penalty-based approaches, and their tuning parameters are fixed (unlike the PX-EM algorithm). They are often used in practice, and their $\mathsf{R}$ packages $\mathsf{elasticnet}$ and $\mathsf{sparsepca}$ are available on CRAN.

To determine the optimal values of the tuning parameters for each algorithm, we consider a vector containing 100 values and estimate the Frobenius loss of the projection matrix for each value in ascending order. The tuning parameter that results in the smallest Frobenius loss value is selected as the optimal value.
The results are presented in Table \ref{sim-3}. Notably, we observed that the batch PX-CAVI algorithm outperforms the other three algorithms listed in the table with the smallest estimation and selection errors, regardless of the values of $p$, $s^\star$, and $r^\star$. Furthermore, the algorithm proposed by \citet{zou06} shows better performance than \citet{erichson20}'s method when $r^\star$ is large. As expected, all three algorithms (batch PX-CAVI and two penalty methods) outperform the conventional PCA method.

\begin{table}[h!]
\begin{adjustwidth}{-.3in}{-.3in}  
\centering
\caption{\small Simulation results of the batch PX-CAVI (bPX-CAVI) algorithm, the two SPCA algorithms proposed by \citet{zou06} and \citet{erichson20} (namely, sPCA1 and sPCA2), and PCA. We fixed $n = 200$ and chose $s^\star \in \{10, 20, 40, 70, 150\}$, $r^\star \in \{1, 2, 3, 5\}$, and $p \in \{500, 1000, 2000, 4000\}$. For each setting, we ran 1000 simulations and obtained the average values of the Frobenius loss of the projection matrix and the percentage of misclassification.}
\label{sim-3}
{ 
\small
\begin{tabular}{ccc|cccc||ccc}
\Xhline{2\arrayrulewidth}
\multicolumn{3}{c|}{ }  & \multicolumn{4}{c||}{\bf Frobenius loss} & \multicolumn{3}{c}{\bf Misclassification (\%)} 
\\\Xhline{2\arrayrulewidth}
    $p$ & $s^\star$ & $r^\star$ &  bPX-CAVI & sPCA1 & sPCA2 & PCA & bPX-CAVI & sPCA1 & sPCA2 \\
\hline
    $1000$ & $10$ & $1$  &
    0.025 & 0.073
    & 0.066 & 0.222 
    & 0.1 & 0.1 & 0.1\\
    $1000$ & $10$ & $3$ &
    0.046 & 0.169 
    & 0.359 & 0.461
    & 0.1 & 0.3 & 10.5 \\
    $1000$ & $10$ & $5$ &
    0.052& 0.207 
    & 0.607  & 0.593
    &0.1 & 0.6 & 26.5\\
    \hline
    $1000$ & $40$ & $1$  &
    0.061 & 0.115 
    & 0.114  & 0.222 
    &0.5 & 0.8 & 0.5\\
    $1000$ & $40$ & $3$ &
    0.131 & 0.261 
    & 0.348 & 0.462 
    & 0.6 & 2.7 & 8.3\\
    $1000$ & $40$ & $5$ &
    0.164 & 0.342 
    & 0.565 & 0.593
    & 0.6 & 2.4 & 26.7\\
    \hline
    $1000$ & $70$ & $1$  &
    0.089 & 0.145 
    & 0.151  & 0.222 
    & 1.2 & 1.8 & 1.2\\
    $1000$ & $70$ & $3$ &
    0.193 & 0.316 
    & 0.368  & 0.462
    & 1.3 & 3.4 & 9.1 \\
    $1000$ & $70$ & $5$ &
    0.247 & 0.417 
    & 0.565  & 0.593 
    & 1.3 & 14.5 & 27.3\\   
    \hline
    $1000$ & $150$ & $1$  &
    0.145 & 0.189 
    & 0.203 & 0.223  
    & 3.5 & 4.2 & 3.5\\
    $1000$ & $150$ & $3$ &
    0.321 & 0.405 
    & 0.422  & 0.463 
    & 4.1 & 8.4 & 12.3 \\
    $1000$ & $150$ & $5$ &
    0.405  & 0.526 
    & 0.594 & 0.593 
    & 4.1 & 16.3 & 28.7\\   
\hline 
    $500$ & $20$ & $2$  &
    0.068 & 0.183 
    & 0.257  & 0.273 
    & 1.3 & 2.3 & 1.2
    \\
    $1000$ & $20$ & $2$ &
    0.068 & 0.175 
    & 0.268 &0.383 
    & 0.1 & 1.8 & 0.1\\
    $2000$ & $20$ & $2$ &
    0.070 & 0.215 
    & 0.285  & 0.532
    & 0.1 & 0.6 & 0.2 \\
    $4000$ & $20$ & $2$  &
    0.072 & 0.332 
    & 0.304  & 0.725
    & 0.1 & 0.1& 0.0 \\
\Xhline{2\arrayrulewidth}    
\end{tabular}  
}
\end{adjustwidth}
\end{table}

\subsection{Unknown $r$}
\label{sim-4}

The previous three studies assumed that $r$ is known. In this study, we investigate the scenario where $r$ is unknown. Rather than modifying our algorithms to estimate $r$ directly—which could increase computation time and introduce complexity in choosing initial values—we propose a practical approach of plugging in a value for $r$ before conducting the analysis. This plugged-in value can be obtained through other algorithms or based on prior studies. Importantly, the accuracy of the plugged-in value is not critical.
This study is designed as follows: We set $r^\star = 4$, $n = 200$, $p = 1000$, and $s^\star = 70$. The input value of $r$ is chosen to be $r = 1, 2, 3, 4, 5, 20$.
For each value, we ran the PX-CAVI algorithm and obtained the average values of $|\langle \widehat  U_{\cdot k}, U_{\cdot k}^\star \rangle|$ and the percentage of misclassification from 1000 simulations.
Note that $\widehat  U_{\cdot k}$ is the $k$-th eigenvector from $\widehat  \mu$; 
$\widehat  U_{\cdot k}$ and $U_{\cdot k}^\star$ are close if $|\langle \widehat  U_{\cdot k}, U_{\cdot k}^\star \rangle|$ is close 1.
The results are provided in Table \ref{sim-4}. From that table, we found that regardless of the input value $r$, even when $r = 20$, the results are similar. Additionally, we noticed that as the rank increases, the accuracy of variable selection improves. 

\begin{table}[h!]
\centering
\caption{\small Simulations for the PX-CAVI algorithm choosing different input values for $r$. Let $r^\star = 4$ , $n = 200$, $p = 1000$, and $s^\star = 70$, and we generated 1000 datasets. For each value $r \in \{1, 2, 3, 4, 5, 20\}$, we calculated the average values (and the standard errors) of the quantity, $|\langle \widehat  U_{\cdot k}, U_{\cdot k}^\star \rangle|$,
and the percentage of misclassification.}
\label{sim-4}
{ \small
\begin{tabular}{cccccc}
\Xhline{2\arrayrulewidth}
   & $|\langle \widehat  U_{\cdot 1}, U_{\cdot 1}^\star \rangle|$ 
   & $|\langle \widehat  U_{\cdot 2}, U_{\cdot 2}^\star \rangle|$ 
   & $|\langle \widehat  U_{\cdot 3}, U_{\cdot 3}^\star \rangle|$ 
   & $|\langle \widehat  U_{\cdot 4}, U_{\cdot 4}^\star \rangle|$ 
   & Misc (\%)\\
\hline
   $r = 1$  & 0.868 (0.180) &  &  & & 1.1 (0.3) \\
   $r = 2$  & 0.864 (0.188) & 0.798 (0.217) & & & 0.2 (0.1) \\
   $r = 3$  & 0.866 (0.186) & 0.801 (0.216) & 0.855 (0.170) & & 0.0 (0.1) \\
   $r = 4$  & 0.869 (0.183) & 0.803 (0.214) & 0.855 (0.171) & 0.932 (0.104) & 0.0 (0.0) \\
   $r = 5$  & 0.868 (0.183) & 0.803 (0.214) & 0.855 (0.170) & 0.933 (0.104) & 0.0 (0.0) \\
   $r = 20$ & 0.881 (0.173) & 0.811 (0.213) & 0.850 (0.176) & 0.933 (0.099) & 0.0 (0.0) \\
\Xhline{2\arrayrulewidth}
\end{tabular}    
}
\end{table}

\section{A real data study}
\label{sec:realstudy}

This section applies the PX-CAVI and batch PX-CAVI algorithms to analyze a lung cancer dataset. The dataset, accessible through the $\mathsf{R}$ package $\mathsf{sparseBC}$, comprises expression levels of 5000 genes and 56 subjects. These subjects encompass 20 pulmonary carcinoid subjects (carcinoid), 13 colon cancer metastasis subjects (colon), 17 normal lung subjects (normal), and 6 small cell lung subjects (small cell). The primary objective is to identify biologically relevant genes correlated with lung cancer and distinguish the four different cancer types.

To prepare the data for analysis, we center and scale it before running each algorithm. In this study, we set the rank $r = 8$, as it captures over $70\%$ variability. Furthermore, we are particularly interested in the first three principal components (PCs). Therefore, selecting $r = 8$ serves the purpose well.
Table \ref{real-1} presents the top 10 reference IDs of genes identified from the first and second PCs. Each reference ID corresponds to a specific gene, and this correspondence can be validated using the NCBI website. For instance, the reference ID `38691\_s\_at' represents the gene 6440 (see \url{https://www.ncbi.nlm.nih.gov/geoprofiles/62830018}).
\begin{table}[h]
\begin{adjustwidth}{-.5in}{-.5in}  
\centering
\caption{\small The results of the top 10 references IDs of genes and the total numbers of active genes of the first two principal components estimated by the PX-CAVI and the batch PX-CAVI (bPX-CAVI) algorithms and PCA.}
\label{real-1}
{ \small
\begin{tabular}{c|ccc||ccc}
\Xhline{2\arrayrulewidth}
 &
\multicolumn{3}{c||}{\bf 1st principal component} & 
\multicolumn{3}{c}{\bf 2nd principal component} \\
\Xhline{2\arrayrulewidth}
Ranking & PX-CAVI & bPX-CAVI & PCA & PX-CAVI & bPX-CAVI & PCA \\
\Xhline{2\arrayrulewidth}
1 & 38691\_s\_at & 38691\_s\_at & 38691\_s\_at & 41209\_at & 41209\_at & 39220\_at\\
2 & 37004\_at & 37004\_at & 37004\_at & 39220\_at & 39220\_at & 41209\_at\\
3 & 33383\_f\_at & 33383\_f\_at & 33383\_f\_at & 38430\_at & 38430\_at & 38430\_at\\
4 & 35926\_s\_at & 35926\_s\_at & 35926\_s\_at & 34708\_at & 34708\_at & 34708\_at\\
5 & 37864\_s\_at & 37864\_s\_at & 37864\_s\_at & 33377\_at & 40607\_at & 40607\_at \\
6 & 41723\_s\_at & 41723\_s\_at & 41723\_s\_at & 40607\_at & 33377\_at & 33377\_at \\
7 & 38096\_f\_at & 38096\_f\_at & 38096\_f\_at & 36780\_at & 36119\_at & 36119\_at\\
8 & 38194\_s\_at & 38194\_s\_at & 38194\_s\_at & 36119\_at & 36780\_at & 36780\_at\\
9 & 33274\_f\_at & 33274\_f\_at & 33274\_f\_at & 32452\_at & 32452\_at & 32452\_at \\
10 & 33500\_i\_at & 33500\_i\_at & 33500\_i\_at & 32052\_at & 35730\_at & 35730\_at\\
\hline
\# of nonzeros & 1183 & 1469 & 5000 & 1183 & 795 & 5000\\
\Xhline{2\arrayrulewidth}
\end{tabular}    
}
\end{adjustwidth}
\end{table}

\begin{figure}[!h]
 \centering
 \caption{The three plots in the first row are the first three principal component scores estimated using the PX-CAVI algorithm and the three plots at the bottom are the same score functions estimated using the batch PX-CAVI algorithm.
}
\label{fig:real-score-fn}
\includegraphics[width=\textwidth]{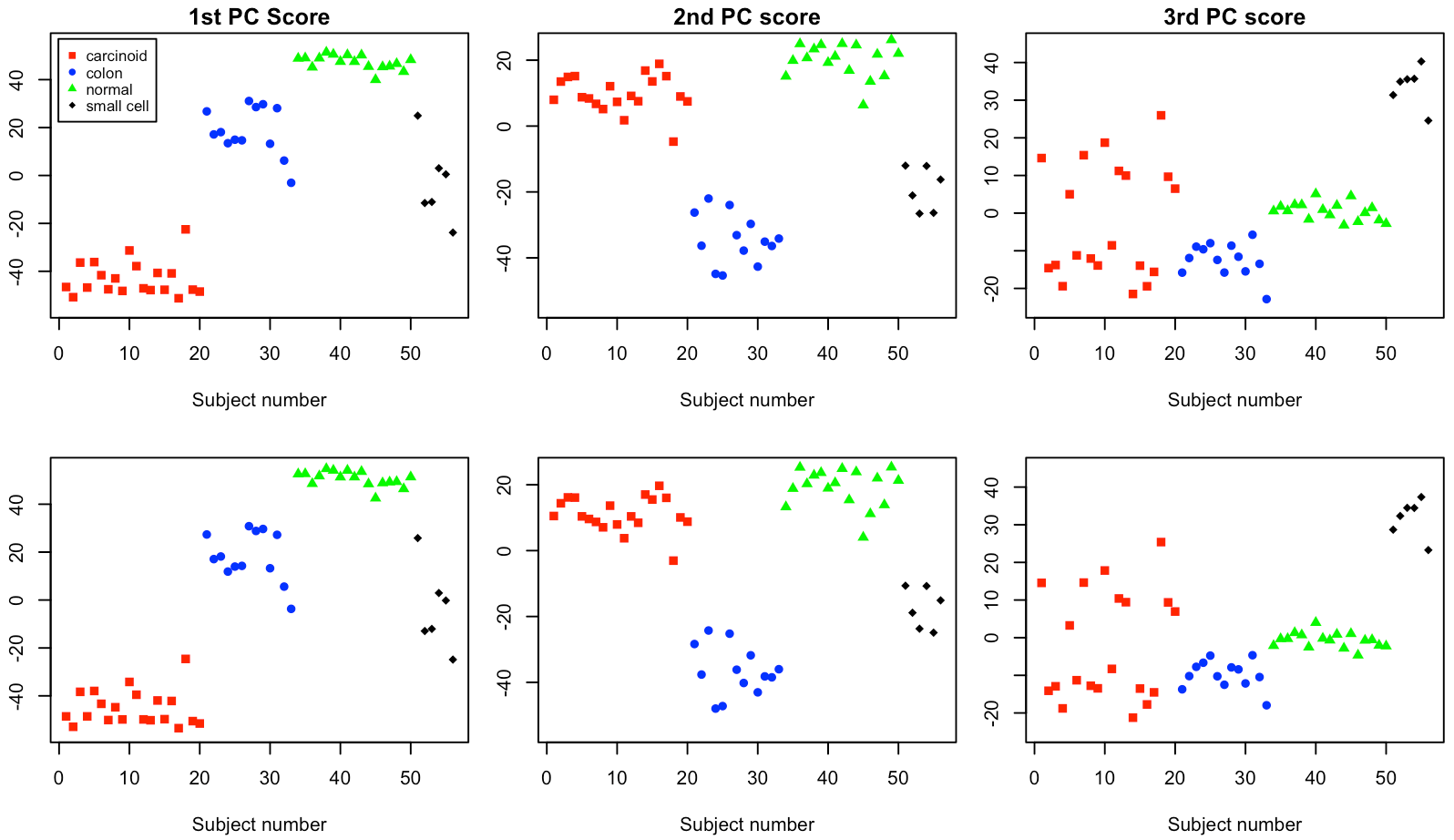}
\end{figure}

From Table \ref{real-1}, we made the following observations.
The top ten genes of the first principal component obtained from all three algorithms are the same. In the second PC, the order might vary slightly, but overall, the results are similar. We conducted a gene count analysis to determine the number of genes with nonzero loading values for each PC. For PCA, which does not impose sparsity on the loadings matrix, the total number of nonzeros is equal to the total number of genes. The PX-CAVI algorithm ensures that all PCs have the same number of nonzero loadings by the jointly row-sparsity assumption. This property leads to easier interpretation, as there is no concern about specific genes being selected in the first PC but not in second PC. The batch PX-CAVI algorithm employs fewer genes than the PX-CAVI algorithm to construct the second PC. By comparing their score functions in Figure \ref{fig:real-score-fn}, we observed that using either 1183 genes or 795 genes to represent PCs does not result in significant differences. This demonstrates the advantage of the batch PX-CAVI algorithm in utilizing fewer genes to construct PCs while maintaining comparable performance.
Additionally, we provided the first three PC scores estimated by both algorithms and highlighted the four different cancer types using different colors. As shown in the PC scores, the four cancer types are well-separated, indicating the effectiveness of our algorithms in distinguishing between the different types of lung cancer.
\section{Conclusion and discussion}
\label{sec:conclusion}

In this paper, we proposed the PX-CAVI algorithm  (also the batch PX-CAVI algorithm) and its EM analogue the PX-EM algorithm for Bayesian SPCA. These algorithms utilized parameter expansion to effectively handle the orthogonality constraint imposed by the loading matrix and enhance their convergence speeds. We demonstrated that the PX-CAVI algorithm outperforms all other algorithms discussed in the paper, showcasing its superiority. Furthermore, we studied the posterior contraction rate of the variational posterior, providing a novel contribution to the existing literature. Additionally, our findings revealed that choosing the normal (or multivariate normal) density for $g$ yielded better results compared to the heavier-tailed Laplace density.

Future studies can delve into understanding why the Laplace density fails to yield smaller estimation errors even in the rank one case. Additionally, the uncertainty quantification problem of SPCA remains unexplored, despite the rich literature on this topic for the sparse linear regression model \citep[see][]{vdP17, belitser20, cast20b, martin20}. Moreover, gaining a deeper understanding of the variational posterior, i.e. its conditions for achieving variable selection consistency would be valuable.
Our $\mathsf{R}$ package $\mathsf{VBsparsePCA}$ for the PX-CAVI and batch PX-CAVI algorithms is available on CRAN, offering a practical tool for researchers to apply these algorithms in their analyses.

\appendix

\section{Derivation of (\ref{eqn:cavi-u})-(\ref{eqn:cavi-sigma-mvn})}
\label{derive-cavi}

First, we need the following result:
\begin{align}
\label{pf-cavi-derv-1}
    \mathbb{E}_{w|\Theta^{(t)}} 
    \left[
        \frac{1}{2\sigma^2} \sum_{i=1}^n (X_{ij} - \widetilde  \beta_j w_i)^2 
    \right]
    = 
    \frac{1}{2\sigma^2}
        \sum_{i=1}^n \left(
            X_{ij}^2 - 2X_{ij} \widetilde  \beta_j \widetilde  \omega_i + \widetilde  \beta_j H_i \widetilde  \beta_j'
        \right),
\end{align}
where 
$H_i = \widetilde  \omega_i\widetilde  \omega_i' + \widetilde  V_w$ and the expressions of
$\widetilde  w_i$ and $\widetilde  V_w$ are given in (\ref{eqn:cavi-w}).

Since the ELBO is a summation of $p$ terms, we solve $u_j$ and $M_j$ for each $j$. 
As the posterior conditional on $\gamma_j =0$ is singular to the Dirac measure, we only need to consider the case $\gamma_j = 1$. This leads to minimize the function
\begin{align*}
    & \mathbb{E}_{\widetilde  u_j, \widetilde M_j, z_j|\gamma_j = 1}\Bigg[
      \frac{1}{2\sigma^2} \sum_{i=1}^n 
      \left(
        - 2X_{ij} \widetilde \beta_j \widetilde  \omega_i + \widetilde \beta_j H_i \widetilde \beta_j' 
      \right)
      + \log \frac{N(\widetilde  u_j, \sigma^2 \widetilde  M_j)}{\kappa_j^\circ g( \widetilde  \beta_j|\lambda_1)}
    \Bigg]\\
    & \quad = C
    - \frac{1}{\sigma^2} \sum_{i=1}^n X_{ij} \widetilde  u_j \widetilde  \omega_i
    + \frac{1}{2\sigma^2} \sum_{i=1}^n \left(
        \widetilde  u_j H_i \widetilde  u_j'
        + \Tr\left(\sigma^2\widetilde  M_jH_i\right)
    \right)\\
    &\quad\quad+ \lambda_1 \sum_{k=1}^r f(\widetilde  u_{jk}, \widetilde  M_{j,kk}),
\end{align*}
where $\kappa_j^\circ = \int \pi(\gamma_j|\kappa) d\Pi(\kappa)$. Then
we take the derivative of $\widetilde  u_j$ and $\widetilde  M_j$ to obtain (\ref{eqn:cavi-u}) and (\ref{eqn:cavi-Xi}).
The solutions in (\ref{eqn:cavi-u-mvn}) are obtained by changing 
$\lambda_1 \sum_{k=1}^r f(\widetilde  u_{jk}, \widetilde  M_{j,kk})$ in the last display with 
$\frac{\lambda_1}{2\sigma^2} \left(\widetilde  u_j \widetilde  u_j' + \sigma^2 \Tr(M_j)\right)$.

To derive (\ref{eqn:cavi-z}), we have  
\begin{align}
&\mathbb{E}_P \left(\mathbb{E}_{w|\Theta^{(t)}} 
\pi(\widetilde  \beta_j, w, X) - \log q(\widetilde  \beta_j)
\right) \nonumber \\
    & = C + \mathbb{E}_{\widetilde  \mu_j, \widetilde  M_j, z_j}
    \Bigg[
     \frac{1}{2\sigma^2} \sum_{i=1}^n 
      \left(
        -2X_{ij} \widetilde  \beta_j \widetilde  \omega_i + \widetilde  \beta_j H_i \widetilde  \beta_j' 
      \right)  + \mathbbm{1}_{\{\gamma_j = 0\}}\log \frac{1 - z_j}{1-\kappa_j^\circ}
     \nonumber\\
    &  \hspace{6.4cm}
    + \mathbbm{1}_{\{\gamma_j = 1\}}\log \frac{z_j N(\widetilde \mu_j, \sigma^2 M_j)}{\kappa_j^\circ g(\widetilde  \beta_j|\lambda_1)} 
    \Bigg]\nonumber\\
    & = C+ (1-z_j) \log \frac{1-z_j}{1-\kappa_j^\circ}\label{eqn:display-1}\\
    &\quad +z_j
    \Bigg\{
    \frac{1}{2\sigma^2} \sum_{i=1}^n 
    \left(
        \widetilde \mu_j H_i \widetilde \mu_j'
        +\sigma^2 \Tr(\widetilde M_j H_i)
        - 2X_{ij} \widetilde \mu_j \widetilde  \omega_i
    \right)+ r\log \left(\frac{\sqrt{2}}{\sqrt{\pi} \sigma \lambda_1}\right)
       \nonumber\\
    & \hspace{3cm}        
      - \frac{1}{2}\log \det(\widetilde M_j) - \frac{1}{2}+ \lambda_1 \sum_{k=1}^r
    f(\widetilde \mu_{jk},\sigma^2 \widetilde M_{j,kk})
    + \log \frac{z_j}{\kappa_j^\circ}
    \Bigg\}.\nonumber
\end{align}

The solution of $\widehat  h_j$ can be obtained by minimizing $z_j$ from the last line of the above display.
Similarly, (\ref{eqn:cavi-z-mvn}) is obtained by minimizing $z_j$ from the following expression 
\begin{equation}
\begin{split}
 & C+z_j\Bigg\{
    \frac{1}{2\sigma^2} \sum_{i=1}^n 
    \left(
        \widetilde \mu_j H_i \widetilde \mu_j'
        +\sigma^2 \Tr(\widetilde M_j H_i)
        - 2X_{ij} \widetilde \mu_j \widetilde  \omega_i
    \right)
     - \frac{r\log \lambda_1 + 1}{2}
    \\
    & \hspace{2.5cm}
      - \frac{1}{2}\log \det(\widetilde M_j) + \frac{\lambda_1}{2\sigma^2}
       \left(
       	 \widetilde  u_j \widetilde  u_j' + \Tr(\sigma^2 \widetilde  M_j)
       \right)
    + \log \frac{z_j}{\kappa_j^\circ}
    \Bigg\}\\
    &\quad+ (1-z_j) \log \frac{1-z_j}{1-\kappa_j^\circ}.
    \label{eqn:display-2}
    \end{split}
\end{equation}

Last, to obtain (\ref{eqn:cavi-sigma}), we first
sum the expressions in (\ref{eqn:display-1}) for all $j =1, \dots, p$. Next, we write down the explicit expression of $C$ which involves $\sigma^2$, i.e.,
$$
pC_{\sigma^2} = \frac{(np + 2\sigma_a +2)\log \sigma^2}{2} + \frac{\Tr(X'X) + 2\sigma_b}{2\sigma^2}.
$$
Last, we plugging the above expression and solve $\sigma^2$. 
The solution (\ref{eqn:cavi-sigma-mvn}) can be obtained similarly using (\ref{eqn:display-2}).

\begin{supplement}
\stitle{Supplement to ``Spike and slab Bayesian sparse principal component analysis''}
\sdescription{In this supplementary material, we present the batch PX-CAVI algorithm, include the simulation results of the the PX-EM algorithm by choosing $\ell_1$-norm and $\ell_2$-norm in its penalty term, give the proofs of Theorems \ref{Thm: Post-cont-rate} and \ref{Thm: variational-post-contr-rate} and Lemma \ref{lemma-convergence-EM}, and provide some auxiliary lemmas.}
\end{supplement}


\section*{Acknowledgements}
We would like to warmly thanks Drs. Ryan Martin and Botond Szab\'o for their helpful suggestions on an early version of this paper. Bo Ning gratefully acknowledges the funding support provided by NASA XRP 80NSSC18K0443.

\bibliographystyle{chicago}
\bibliography{citation.bib}

\end{document}


\begin{frontmatter}

\title{Supplement to ``Spike and slab Bayesian sparse principal component analysis''}
\runtitle{Spike and slab Bayesian SPCA}

\author{\fnms{Bo Y.-C.} \snm{Ning}\ead[label=e1]{yuchien.ning@gmail.com}}
\and
\author{\fnms{Ning} \snm{Ning}\ead[label=e2]{patning@tamu.edu}}

\affiliation{Harvard University\thanksmark{m1}}

\address{Harvard T. H. Chan School of Public Health \\ 
	Department of Epidemiology, \\
	677 Huntington Ave, Boston, MA 02115\\
	\printead{e1}}
\address{Department of Statistics\\
	Texas A\&M University\\
	College Station, TX 77843
	\printead{e2}}

\runauthor{Ning and Ning}

\end{frontmatter}


This supplementary material contains four sections. Section \ref{app:batch-PX-CAVI} provides the batch PX-CAVI algorithm. Section \ref{px-em} gives the simulation results of the PX-EM algorithm using the $\ell_1$-norm and $\ell_2$-norm in the penalty function. Section \ref{sec:proofs} gives the proofs of Theorems \ref{Thm: Post-cont-rate} and \ref{Thm: variational-post-contr-rate} and Lemma \ref{lemma-convergence-EM}. Auxiliary lemmas are presented in Section \ref{sec:aux}.

\section{The batch PX-CAVI algorithm}
\label{app:batch-PX-CAVI}

In this section, we derive the batch PX-CAVI algorithm. Unlike the PX-CAVI algorithm, the batch PX-CAVI algorithm does not assume $\theta$ being jointly row-sparse; each column of $\theta$ can have identical support such that $S_k \neq S_{k'}$, where $S_k = \{j \in \{1, \dots, p\}: \theta_{jk} \neq 0\}$. 
We need to modify the spike and slab prior, which is
\begin{align}
\pi(\theta, \bgamma|\lambda_1, r) 
& \propto
\prod_{j=1}^p 
\Bigg\{
\int_{A \in V_{r, r}} 
\prod_{k=1}^r
\Big[
\gamma_{jk} g(\theta_{jk}|\lambda_1, A, r) + (1-\gamma_{jk}) \delta_0(\theta_{jk})
\Big] \pi(A) dA
\Bigg\}, \\
&\gamma_{jk}|\kappa \sim \text{Bernoulli}(\kappa), \quad\text{and}\quad \kappa \sim \text{Beta}(\alpha_1, \alpha_2), \nonumber 
\end{align}
where $\gamma_{jk} \in \{0, 1\}$. 
The mean-field variational class is given by
\begin{equation}
\begin{split}
\widetilde{\mathcal{P}}^{MF} = \Bigg\{
& P(\theta): \prod_{j=1}^p \prod_{k=1}^r \Big[z_{jk} \mathcal{N}(\mu_{jk}, \sigma^2 \sigma_{jk}^2)
+ (1-z_{jk}) \delta_0\Big],\; \mu_{jk} \in \mathbb{R},\\
&\hspace{2cm} \langle \mu_{\cdot k}, \mu_{\cdot k'} \rangle = 0, \;\forall k \neq k', \;
\sigma_{jk} \in \mathbb{R}^+, \;z_{jk} \in [0, 1]
\Bigg\}.
\end{split}
\end{equation}
Then the mean-field variational posterior becomes
$$
\widehat P(\theta) = \argmin_{P(\theta) \in \widetilde{\mathcal{P}}^{MF}} KL(P(\theta), \pi(\theta|X)).
$$
Similar to the PX-CAVI algorithm, the batch PX-CAVI algorithm includes an E-step. 
It maximizes the objective function, which is 
$$
\sum_{j=1}^p \sum_{k=1}^r \left(
\mathbb{E}_P Q(\Theta_{jk}|\Theta^{(t)}) - \mathbb{E}_P P(\theta)
\right),
$$
where $Q(\Theta|\Theta^{(t)}) = \mathbb{E}_{\bw|\Theta^{(t)}} \log (\theta, \bw, X)$ and 
$\Theta = (\theta, \Psi, \bz)$ with $\Psi$ given below.
We also apply parameter expansion twice to the likelihood and let $g$ be the multivariate normal density.

In each iteration, the batch PX-CAVI algorithm updates $(\theta_{j1}, \dots, \theta_{jr})$ simultaneously for each $j$. 
We define 
\begin{equation*}
\Psi = 
\begin{pmatrix}
\sigma_{11}^2 & \cdots & \sigma_{1r}^2 \\
\vdots & \vdots & \vdots  \\
\sigma_{p1}^2 & \cdots & \sigma_{pr}^2
\end{pmatrix}
\end{equation*}
as the variance matrix for $P(\theta)$ and denote $\widetilde\Psi$ as the variance matrix for $q(\widetilde \beta)$.
Its $j$-th row is denoted by $\Psi_{j} = (\sigma_{j1}^2, \cdots, \sigma_{jr}^2)'$ and similarly, $\widetilde \Psi_j = (\widetilde \sigma_{j1}^2, \dots, \widetilde \sigma_{jr}^2)$.
Then we obtain
\begin{align}
\label{bpx-cavi:u}
\widehat{\widetilde u}_j &= \left(
\lambda_1 I_r + \sum_{i=1}^n H_i
\right)^{-1} \left(\sum_{i=1}^n X_{ij} \widetilde \omega_i\right), \\
\widehat {\widetilde \Psi}_{j} &= \text{Diag}
\left(
\lambda_1 I_r + \sum_{i=1}^n H_i
\right)^{-1},
\label{bpx-cavi:psi}
\end{align}
where $\text{Diag}(A)$ is a vector containing the diagonal elements of the matrix $A \in \mathbb{R}^{r \times r}$.
Recall that $H_i = \widetilde \omega_i \widetilde \omega_i' + \widetilde V_w$.
Denote $a \circ b$ as the point-wise product between vectors $a$ and $b$, we also obtain 
\begin{align}
\widehat h_j  
\label{bpx-cavi:h}
& = - \frac{1}{2\sigma^2} \sum_{i=1}^n 
\left(
- 2 X_{ij} \widetilde u_j' \circ \widetilde w_i + \text{Diag}(\widetilde u_j' \widetilde u_j H_i) + \sigma^2 \widetilde \Psi_j \circ \text{Diag}H_i 
\right)\nonumber \\
& \quad + \left[\log\left(\frac{\alpha_1}{\alpha_2}\right) + \frac{1}{2} + \frac{\log \lambda_1}{2}\right] \mathbbm{1}_r + \frac{\log (\widetilde\Psi_j)}{2}
-\frac{\lambda_1}{2} \left(\widetilde u_j' \circ \widetilde u_j' + \sigma^2 \widetilde \Psi_j\right).
\end{align}

Once we obtained $\widehat{\widetilde u}$ and $\widehat {\widetilde \Psi}$, we can also obtain $\widehat \mu$ and $\widehat \Psi$ (the same as that in the PX-CAVI algorithm). 
Since batch PX-CAVI does not use jointly row-sparsity, the support of $\widetilde \beta$ is different from the support of $\theta$. The $\widehat { h}$ obtained above is for $\widetilde \beta$. To obtain it for $\theta$, say $\widehat { h}_\theta$, we need to plug-in the estimated values $\widehat \theta$ and $\widehat \Psi$ into (\ref{bpx-cavi:h}). We then solve $\widehat \bz_{\theta}$ from $\widehat h_{jk, \theta} = \log (\widehat z_{jk, \theta} / (1- \widehat z_{jk, \theta})$.

Last, denote $m_j = z_j \circ \widetilde u_j$, we obtain
\begin{align}
\label{bpx-cavi:sigma}
\widehat \sigma^2 = 
\frac{
\Tr(X'X) + \sum_{j=1}^p \sum_{i=1}^n 
\left(
m_j H_i m_j'
- 2X_{ij} m_j \widetilde w_j
+ \lambda_1 m_j m_j' 
\right)+ 2\sigma_b
}{
np + 2(\sigma_a + 1)
}.
\end{align}
The batch PX-CAVI algorithm is given below.
\begin{algorithm}[ht]
\DontPrintSemicolon
   \KwData{$X$, a $p \times n$ matrix, centered and scaled}
   
  \KwInput{$\widehat \mu^{(0)}$, $\widehat{\Psi}^{(0)}$, $\widehat\bz^{(0)}$, $\widehat\sigma^{(0)}$, $r$, number of total iterations $T$, and the threshold $\delta$}
  
{\bf For $t = 0, \dots, T-1$, repeat}:

  \begin{itemize}
  \item[--] update $\widetilde \bomega^{(t+1)}$ and $\widetilde V_w^{(t+1)}$ from (\ref{eqn:cavi-w})
  \item[--] update ${\widetilde u}^{(t+1)}$ and ${\widetilde \Psi}^{(t+1)}$ from (\ref{bpx-cavi:u}) and (\ref{bpx-cavi:psi}) 
  \item[--] update ${ h}^{(t+1)}$ from (\ref{bpx-cavi:h}), then, obtain 
  $\widehat \bz^{(t+1)}$
  \item[--] update $\sigma^{(t+1)}$ from (\ref{bpx-cavi:sigma}) 
    \item[--] obtain $D^{(t+1)}$, $u^{(t+1)}$, and $\Psi^{(t+1)}$
  \item[--] apply SVD to obtain $A^{(t+1)}$ and then, obtain $\mu^{(t+1)}$
  \item[--] using $\mu^{(t+1)}$ and $\Psi^{(t+1)}$ to obtain ${ h}_{\theta}$ and $\bz_\theta$
\end{itemize}

{\bf Stop}: if $\max\left(\big\|\mu^{(t+1)} \mu^{{(t+1)}'} - \mu^{(t)} \mu^{{(t)}'}\big\|_F, \|  \bz_\theta^{(t+1)} - \bz_\theta^{(t)}\|_1\right) \leq \delta$

 \KwOutput{$\widehat P(\theta)$. }
\caption{The batch PX-CAVI algorithm}
\label{batch-px-cavi}
\end{algorithm}

\section{The PX-EM algorithm: $\ell_1$-norm versus $\ell_2$-norm}
\label{px-em}

We conduct a simulation study to compare the use of the $\ell_1$- and $\ell_2$-norms of the penalty function of the PX-EM algorithm in Section \ref{sec:EM}. 
We chose $n = 200$, $s^\star \in \{10, 20, 40, 70, 150\}$, $r^\star \in \{1, 2, 3, 5\}$, and $p \in \{500, 1000, 2000, 4000\}$. We run 100 simulations in each setting.
From Table \ref{sim-5}, we observe that the $\ell_1$-norm gives more accurate results on both parameter estimation and variable selection, particularly when $r^\star$ is large. 

\begin{table}[!]
\begin{adjustwidth}{-.5in}{-.5in}  
\centering
\caption{\small Simulation results of the PX-EM algorithm using $\ell_1$- and $\ell_2$-norm. We chose $n = 200$, $s^\star \in \{10, 20, 40, 70, 150\}$, $r^\star \in \{1, 2, 3, 5\}$, and $p \in \{500, 1000, 2000, 4000\}$. For each setting, we ran $100$ simulations and obtained the average values of the Frobenius loss of the projection matrix, the percentage of misclassification, FDR, and FNR.}
\label{sim-5}
{ 
\small
\begin{tabular}{ccc|cc|cc|cc|cc}
\Xhline{2\arrayrulewidth}
\multicolumn{3}{c|}{} & \multicolumn{2}{c|}{\bf Frobenius loss} & \multicolumn{2}{c|}{\bf Misc (\%)}
& \multicolumn{2}{c|}{\bf FDR} & \multicolumn{2}{c}{\bf FNR}
\\\Xhline{2\arrayrulewidth}
    $p$ & $s^\star$ & $r^\star$ &  $\ell_1$-norm & $\ell_2$-norm
    &  $\ell_1$-norm & $\ell_2$-norm &  $\ell_1$-norm & $\ell_2$-norm &  $\ell_1$-norm & $\ell_2$-norm\\
\hline
    $1000$ & $10$ & $1$  &
    0.024 & 0.026 &
    0.1 &  0.1 &
    0.001 & 0.000 & 0.001 & 0.001\\
    $1000$ & $10$ & $3$ &
    0.040 & 0.041  &
    0.0 &  0.0 &
    0.000 & 0.000 & 0.000 & 0.000\\
    $1000$ & $10$ & $5$ &
    0.043 & 0.045 &
    0.0 & 0.0 
    & 0.000 & 0.000 & 0.000 & 0.000\\ 
    \hline
    $1000$ & $40$ & $1$  &
    0.054  & 0.054  &
    0.4 & 0.4  &
    0.001 & 0.001 & 0.004 & 0.004\\
    $1000$ & $40$ & $3$ &
    0.128 & 0.186 &
    0.1 &  0.2
    & 0.000 & 0.000 & 0.001 & 0.002\\
    $1000$ & $40$ & $5$ &
    0.128 & 0.241 &
    0.1 &  0.2 & 
    0.000 & 0.000 & 0.001 & 0.002\\
    \hline
    $1000$ & $70$ & $1$  &
    0.093 & 0.092  &
    1.2 & 1.2 &
    0.000 & 0.000 & 0.013 & 0.013\\
    $1000$ & $70$ & $3$ &
    0.214 & 0.335  &
    0.4 &  0.9  &
    0.000 & 0.000 & 0.005 & 0.009\\
    $1000$ & $70$ & $5$ &
    0.212 & 0.514 &
    0.1 &  0.7
    &0.000 & 0.000 & 0.002 & 0.008\\   
    \hline
    $1000$ & $150$ & $1$  &
    0.155 & 0.155 &
    3.7 &  3.7
    & 0.000  & 0.000 & 0.042 & 0.042\\
    $1000$ & $150$ & $3$ &
    0.463 & 0.763 &
    2.6 &  4.8
    & 0.000 & 0.000 & 0.029 & 0.054\\
    $1000$ & $150$ & $5$ &
    0.520 & 1.363 &
    1.5 &  5.9 &
    0.000 & 0.000 & 0.017 & 0.065\\   
\Xhline{2\arrayrulewidth}    
\end{tabular}  
}
\end{adjustwidth}
\end{table}
\FloatBarrier

\section{Proofs of Theorems \ref{Thm: Post-cont-rate} and \ref{Thm: variational-post-contr-rate} and Lemma \ref{lemma-convergence-EM}}
\label{sec:proofs}

\subsection{Proof of Theorem \ref{Thm: Post-cont-rate}}

\begin{lemma}
\label{lemma-1}
For some sufficiently large $C_1$, 
under Assumption \ref{assumps},
we have that as $n$ goes to infinity,
$$\displaystyle
\mathbb{P}_{\Sigma^\star}
\left( 
\int f/f^\star d \Pi(\theta) \leq \exp(-C_1 n\epsilon_n^2)
\right)
\to 0.
$$
\end{lemma}
\begin{proof}
By Lemma 8.10 of \citet{ghosal17}, it suffices to show that
\begin{align}
\label{pf-post-contr-rate-1}
    \Pi \Big(K(f^\star, f) \leq n\epsilon_n^2, \;
    V(f^\star, f) \leq n\epsilon_n^2 \Big) 
    \gtrsim \exp(-C_1 n\epsilon_n^2),
\end{align}
where $V(f^\star, f)$ is the Kullback-Leibler variation between $f^\star$ and $f$.
Since both $f = \prod_{i=1}^n f_i$ and $f^\star = \prod_{i=1}^n f^\star_i$ are independent multivariate Gaussian densities, we have
\begin{align*}
    \frac{1}{n}K(f^\star, f) & =  
    \frac{1}{n} \sum_{i=1}^n K(f_i^\star, f_i)
    = \frac{1}{2}\Big[
    \Tr(\Sigma^{-1} \Sigma^\star) - p - \log(\det(\Sigma^{-1}\Sigma^\star))
    \Big],\\
    \frac{1}{n}V(f^\star, f) & =  
    \frac{1}{n} \sum_{i=1}^n V(f_i^\star, f_i)
    = 
    \frac{1}{2}
    \Big[
    \Tr(\Sigma^{-1} \Sigma^\star\Sigma^{-1} \Sigma^\star)  
    - 2\Tr(\Sigma^{-1} \Sigma^\star) + p)
    \Big].
\end{align*}
Let $\tSigma = {\Sigma^\star}^{1/2} \Sigma^{-1} {\Sigma^\star}^{1/2}$ and $\rho_j$ be the $j$-th largest eigenvalue of $\tSigma$,
we have
\begin{align}
    & \hspace{-0.5cm}\Pi
   \Big(K(f^\star, f) \leq n\epsilon_n^2, \
    V(f^\star, f) \leq n\epsilon_n^2
    \Big) \nonumber \\
    & \geq 
    \Pi \left(
    \sum_{j=1}^p (\rho_j - 1 - \log \rho_j) \leq 2\epsilon_n^2, \
    \sum_{j=1}^p (\rho_j - 1)^2 \leq 2\epsilon_n^2
    \right)
    \nonumber \\
    & \geq
    \Pi \left(
    \sum_{j=1}^p (\rho_j - 1)^2 \leq 2\epsilon_n^2
    \right)\\
    &\geq 
    \Pi\left(
    	\|\Sigma^{-1}\| \|\Sigma^\star - \Sigma\|_F \leq \sqrt{2} \epsilon_n
    \right) 
    \nonumber \\
    &  =
    \Pi\left(
        \|\Sigma - \Sigma^\star\|_F \leq \sqrt{2} \sigma^2 \epsilon_n
    \right)\\
    &= \Pi\left(
        \|\beta\beta' - \beta^\star {\beta^\star}'\|_F \leq \sqrt{2}\sigma^2 \epsilon_n
    \right).
    \label{pf-post-contr-rate-1.1}
\end{align}
In the last display, the second lower bound is obtained by applying the inequality $\log(1+x) \geq x - x^2/2$ for $x < 1$.
The last inequality is obtained by using the fact that
$\|\Sigma^{-1}\| \geq 1/\sigma^2$, and 
the last equality is obtained by $\beta \beta' = \theta \theta'$.
(\ref{pf-post-contr-rate-1.1}) can be further bounded in below by
\begin{align}
    & \Pi\left(
        \|\beta-\beta^\star\|_F(\|\beta\| + \|\beta^\star\|)
        \leq \sqrt{2}\sigma^2 \epsilon_n
    \right) \nonumber \\
   & \quad  \geq 
     \Pi\left(
        \|\beta-\beta^\star\|_F(\|\beta - \beta^\star\|_F + 2\|\beta^\star\|)
        \leq \sqrt{2}\sigma^2\epsilon_n
    \right) \nonumber \\
    & \quad \geq 
    \Pi\left(
    	\|\beta^\star\|\|\beta - \beta^\star\|_F \leq \frac{\sqrt{2}\sigma^2\epsilon_n}{4}
    \right) \nonumber \\
    & \quad \geq
    \Pi\left(
    \|\beta - \beta^\star\|_F 
    \leq \frac{\sqrt{2}\sigma^2 \epsilon_n}{4 \|\beta^\star\|_{q,1}}
    \Bigg| r = r^\star 
\right) \Pi(r = r^\star).
    \label{pf-post-contr-rate-1.3}
\end{align}
We use the inequality $\|\beta - \beta^\star\|_F \leq 2 \|\beta^\star\|$ to obtain second lower bound in the last display. Otherwise, if $2\|\beta^\star\| < \|\beta -\beta^\star\|_F$, then $\|\theta^\star\| \to 0$.
Next, we shall bound (\ref{pf-post-contr-rate-1.3}) below.
Under Assumption \ref{assumps}, $\Pi(r = r^\star) \geq \exp(-a_3 r^\star)$.
Let $d_n = \sqrt{2}\sigma^2 \epsilon_n/(8 \|\beta^\star\|_{q,1})$, we have 
\begin{align}
	& \Pi\left(\|\beta - \beta^\star\|_F \leq 2d_n
    \big| r = r^\star
    \right)\nonumber\\
    &\geq \Pi\left(\sum_{j=1}^p \|\beta_j - \beta^\star_j\|_2 \leq 2d_n
    \Big| r = r^\star
    \right)
    \nonumber \\
    &\geq 
 \frac{\pi(s^\star)}{{p \choose s^\star}}
 \int_A
 \int_{\sum_{j \in S^\star} \|\beta_j - \beta^\star_j\|_q \leq d_n} \prod_{j \in S^\star} g(\theta_j|\lambda_1, A) d\theta_j d\Pi(A)
 \prod_{j \not\in S^\star} \delta_0(\theta_j).
 \label{pf-post-contr-rate-1.4}
\end{align}
Under Assumption \ref{assumps} and using the Stirling approximation to a binomial coefficient, we obtain
\begin{align*}
    \frac{\pi(s^\star)}{{p \choose s^\star}}
    \gtrsim p^{-a_1 s^\star} \left(\frac{p}{s^\star} \right)^{s^\star} 
    \geq \exp \left(-(a_1-1) s^\star \log p - s^\star \log s^\star \right).
\end{align*}
Denote $\check{\beta}_j = \beta_j - \beta^\star_j$ and change the variable from $\beta_j$ to $\check{\beta}_j$,
then the integral in (\ref{pf-post-contr-rate-1.4}) is bounded below by 
\begin{align}
\label{pf-post-contr-rate-1.4.1}
     \exp\left(-\lambda_1 \|\beta^\star\|_{q,1}^m \right)
    \int_{\sum_{j \in S^\star}
    \|\check{\beta}_j \|_q^m \leq d_n}
    C(\lambda_1)^{r^\star}
    \exp
    \left(
        -\lambda_1 \sum_{j \in S^\star} \|\check\beta_j\|_q^m
    \right) d \check\beta_j.
 \end{align}   
 If $m = 1, 1 \leq q \leq 2$, then
 $C(\lambda_1) = \lambda_1/a_{r^\star}$ and
 (\ref{pf-post-contr-rate-1.4.1}) can be bounded by 
 \begin{align}
    & \exp\left(-\lambda_1 \|\beta^\star\|_{q,1}, \right)
    \left(\frac{2}{a_{r^\star}}\right)^{r^\star s^\star} 
    \prod_{j\in S^\star}
    \int_{\sum_{j \in S^\star} 
        \|\check{\beta}_j\|_1 \leq d_n}
    \left(\frac{\lambda_1}{2}\right)^{r^\star} 
    \exp(-\lambda_1 \|\check\beta_j\|_1) d \check\beta_j 
    \nonumber \\
    & \quad \geq 
    \exp\left(
    -\lambda_1 \|\beta^\star\|_{q,1} 
    - \lambda_1 d_n
    \right)
    \left(\frac{2}{a_{r^\star}}\right)^{r^\star s^\star} 
     (\lambda_1 d_n)^{r^\star s^\star} \frac{1}{(r^\star s^\star)!}.
     \label{pf-post-contr-rate-1.5}
\end{align}
Then we plug-in the lower bound of $\|\theta^\star\|$, the upper bound of $\|\theta^\star\|_{1,1}$ (note that $\|\beta^\star\|_{1,1} \leq \|\theta^\star\|_{1,1}$), and the upper bound of $\lambda_1$
to obtain that $$\exp(-\lambda_1 \|\beta^\star\|_{q,1} - \lambda_1 d_n) \geq \exp(-c_{11} n\epsilon_n^2),$$
where $c_{11} = b_3(1+ \sqrt{2}\sigma^2/(8b_4))$. 
Next, since $1 \leq a_r \leq O(\sqrt{r})$, we have that $$(2/a_{r^\star})^{r^\star s^\star} \geq \exp(-c_{12}r^\star s^\star \log r^\star)\geq \exp(-c_{12}s^\star \log p)$$ for some positive constant $c_{12}$.
Furthermore, 
we plug-in the lower bound of $\lambda_1$ and the upper bounds of $\|\theta^\star\|_{q,1}$ and $r^\star$ and using the fact that $s^\star \log p < n$ to obtain that
\begin{align*}
(\lambda_1 d_n)^{r^\star s^\star} & = 
\exp\left(-
r^\star s^\star \log \left(
\frac{8 \|\beta^\star\|_{q,1}}{\sqrt{2}\sigma^2 \lambda_1 \epsilon_n}
\right)
\right)\\
& \geq
\exp\left(- 
r^\star s^\star \log \left(
\frac{8 b_5 s^\star \log p}{\sqrt{2} \sigma^2 \lambda_1^2 \epsilon_n}
\right) \right)\\
& \geq 
\exp\left(-
r^\star s^\star \log \left(
\frac{8 b_5 p^{b_2/r^\star}}{\sqrt{2} \sigma^2 b_1^2}
\right)
\right)\\
& \geq 
\exp(- c_{13} n\epsilon_n^2),
\end{align*}
where $c_{13} = b_2 + \log(8b_5/(\sqrt{2}b_1^2\sigma^2))$.
Last, applying the Stirling's approximation to $(r^\star s^\star)!$ yields $$\exp(-r^\star s^\star \log (r^\star s^\star)) \geq \exp(-s^\star \log p) = \exp(-n\epsilon_n^2).$$
By combining all the lower bounds derived above, we obtain that
\begin{align*}
	& \exp\left(-\lambda_1 \|\beta^\star\|_{q,1}, \right)
	\left(\frac{2}{a_{r^\star}}\right)^{r^\star s^\star} 
	\prod_{j\in S^\star}
	\int_{\sum_{j \in S^\star} 
		\|\check{\beta}_j\|_1 \leq d_n}
	\left(\frac{\lambda_1}{2}\right)^{r^\star} 
	\exp(-\lambda_1 \|\check\beta_j\|_1) d \check\beta_j 
	\nonumber \\
	& \quad \geq \exp(- (c_{11} + c_{12} + c_{13}) n\epsilon_n^2).
\end{align*}
Therefore, we obtain (\ref{pf-post-contr-rate-1}) with $C_1 \geq a_1 + c_{11} + c_{12} + c_{13}$.

If $m = 2$ and $q = 2$, then the term in (\ref{pf-post-contr-rate-1.4.1}) can be bounded in below by 
\begin{align*}
& \exp(-\lambda_1 \|\beta^\star\|^2) \int_{\sum_{j\in S^\star} \|\check{\beta}\|^2 \leq d_n}
\left(\frac{\lambda_1}{2\pi}\right)^{r^\star s^\star/2} 
\exp\Big( - \lambda_1 \sum_{j\in S^\star} \|\check\beta_j\|^2 \Big) d\check\beta \\
& \quad \geq \exp(-\lambda_1 \|\beta^\star\|^2) 
\left(1 - \Phi(|\check\beta_{jk}| \geq \sqrt{\lambda_1 d_n/(s^\star r^\star)})\right)^{r^\star s^\star}\\
& \quad \geq \exp(-\lambda_1 \|\theta^\star\|^2) \left(1 - 2e^{-d_n\lambda_1 / (2s^\star r^\star)}\right)^{s^\star r^\star}\\
& \quad \geq \exp\Big(- \lambda_1\|\theta^\star\|^2 + r^\star s^\star \log (\lambda_1 d_n) - r^\star s^\star \log (r^\star s^\star /8)\Big) \\
& \quad \geq \exp(-c_{14}n\epsilon_n^2),
\end{align*}
where $c_{14} = b_3 b_5 + b_2 + \log(8b_5/(\sqrt{2}b_1\sigma^2)) + 1$.
The second inequality in the last display is obtained by applying the tail bound of the standard normal distribution and using the fact that $\|\beta^\star\| = \|\theta^\star\|$. The third inequality is obtained by using the inequality $1 - e^{-x} \geq x/2$ for $x < 1$ and note that 
\begin{align*}
\exp(-r^\star s^\star \log(\lambda_1 d_n)) &\geq \exp\left(-r^\star s^\star \log \left(\frac{8\|\beta^\star\|^2}{\lambda_1 \sqrt{2}\sigma^2 \epsilon_n}\right) \right)\\
& \geq \exp\left(-s^\star r^\star \log \left(
\frac{8 \sqrt{s^\star \log p} p^{b_2/r^\star}}{\sqrt{2} \sigma^2 n}
\right) \right) \\
& \geq \exp\Big(-(b_2 + \log(8b_5/(\sqrt{2}b_1\sigma^2)) + 1) n\epsilon_n^2\Big),
\end{align*}
where we plugged-in the lower bound of $\lambda_1$ and upper bound of $\|\theta^\star\|^2$ to obtain the second lower bound in the last display.
\end{proof}

\begin{proof}[Proof of Theorem \ref{Thm: Post-cont-rate}]
We first prove (\ref{Thm-1:eqn-3}).
By Lemma \ref{lemma-1}, we only need to take care of the numerator.
Define the set $\mathcal{S} = \{\theta: s \leq M_2s^\star\}$,
we have
\begin{align*}
    \mathbb{E}_{f^\star} \left(
        \int_{\mathcal{S}^c} \frac{f}{f^\star} d\Pi(\theta)
    \right) 
    & \leq \int_{\mathcal{S}^c} d\Pi(\theta)
    \leq \Pi(\theta: s > M_2s^\star)
    = \sum_{s = M_2s^\star}^\infty \pi(s). 
\end{align*}
Under Assumption \ref{assumps},
without loss of generality assuming that $\pi(s)/\pi(s-1) \leq p^{- a_2}$, we obtain that
\begin{align*}
    \sum_{s = M_2s^\star}^\infty \pi(s)
    & \leq 
    \sum_{s = M_2 s^\star}^\infty \pi(s^\star) 
    (p^{-a_2})^{M_2s^\star - s^\star}
    \leq 
    2 p^{-a_2(M_2s^\star - s^\star)}.
\end{align*}
The last inequality is obtained using the trivial bounds
$p^{-a_2} \leq 1/2$ and $\pi(s^\star) < 1$.
Let $M_2 \geq C_1/a_2$, 
we obtain (\ref{Thm-1:eqn-3}).

Now, we proof (\ref{Thm-1:eqn-1}). 
We apply the general theory of posterior contraction \citep{ghosal17}.
The prior mass condition is given in Lemma \ref{lemma-1}.
Next, we need to show that there exists a sieve $\mathcal{F}_n$ such that $\Pi(\mathcal{F}^c_n) \lesssim \exp(-C_2n\epsilon_n^2)$. Consider the following sieve:
$$
\mathcal{F}_n = \left\{\beta = \theta O: \max_j \|\beta_j\|_{q} \leq H_n,\; r \leq J_n\right\},
$$
where $J_n = s^\star \log p$, and $H_n = c n/\left(\underline{\lambda_1}\right)$ with $\underline{\lambda_1}$ is the lower bound of $\lambda_1$ (see Assumption \ref{assumps}) and $c$ is a positive constant.
We shall show that $\Pi(\mathcal{F}_n^c) \lesssim \exp(-C_2 n\epsilon_n^2)$.
Note that
\begin{align}
\label{pf-post-contr-rate-1.6}
\Pi(\mathcal{F}_n^c) \leq 
\Pi\Big(\max_j\|\beta_j\|_q \geq H_n| s \leq M_2s^\star,\; r \leq J_n\Big) + \Pi(r > J_n).
\end{align}
Under Assumption \ref{assumps}, 
we have
\begin{align*}
    \Pi(r > s^\star \log p)
    \lesssim \sum_{r = s^\star \log p}^{\infty} 
    \exp(- a_4 r)
    &\leq \exp(- a_4 s^\star \log p) \sum_{k = 0}^\infty e^{- a_4 k}\\
    &= \exp(- a_4 s^\star \log p)/a_4.
\end{align*}
The first term in the right hand side of (\ref{pf-post-contr-rate-1.6}) is bounded by
\begin{align*}
& \sum_{r = 1}^{J_n}
\sum_{\{S: s \leq M_2s^\star\}} 
   \frac{\pi(s)}{{p \choose s}} 
   \sum_{j \in S} \int_A \Pi(\|\beta_j\|_q \geq H_n|\lambda_1) d\Pi(A) \\
   &\quad \leq 
    \sum_{s = 0}^{M_2s^\star} \pi(s)
    \left(
   \sum_{r = 1}^{J_n}
    M_2s^\star \exp((r - H_n\lambda_1)/2)
   \right)\\
   & \quad \leq
   \sum_{r = 1}^{J_n}
   \Big(
   \exp(\log(M_2s^\star) + (r - cn)/2)
   \Big)\\
   & \quad \leq 
   \exp\Big(\log(M_2s^\star) + \log J_n + J_n/2 - cn/2\Big)\\
 & \quad \leq \exp(- C_2 n \epsilon_n^2),
\end{align*}
where the second line of the last display is obtained by the fact that $\pi(s) < 1$.
Combining the two upper bounds derived above, we obtain that 
$\Pi(\mathcal{F}_n^c) \lesssim \exp(-C_2 n \epsilon_n^2)$,
where $C_2 > C_1 + 4$ and $c > C_1 + 7$.

At last, we need to find a sequence of test $\phi_n$ and show that
\begin{align}
\label{test}
H_0: \mathbb{E}_{f^\star} \phi_n \to 0, \quad
H_1: \sup_{\Sigma \in \mathcal{F}_n: \|\Sigma - \Sigma^\star\| \geq M_2\epsilon_n} \mathbb{E}_{f}(1-\phi_n) \leq \exp(-C_3 n\epsilon_n^2).
\end{align}
We directly use the test constructed by \citet{gao15} (see Lemma \ref{apx-test}) and obtain that 
\begin{align}
\mathbb{E}_{f^\star} \phi_n
& \leq \exp
\left(
C_3 M_2 s^\star - \frac{C_3 M_2^2 n\epsilon_n^2}{4\|\beta^\star\|^2}
\right)
+ 2\exp \left(C_3 M_2 s^\star - C_3 \sqrt{M_2} n \right)
\nonumber \\
& \leq 3\exp
\left(
C_3 M_2 s^\star - \frac{C_3 M_2^2 n\epsilon_n^2}{4b_4^2}
\right).
\label{pf-post-contr-rate-1.7}
\end{align}
The second upper bound in the last display is obtained using the assumption that $\|\beta^\star\| = \|\theta^\star\| \geq b_4$ in Assumption \ref{assump:truevalue}.

To prove the alternative part of the test in (\ref{test}), we divide it into small pieces based on $S$ (recall that $S \subset \{S:s \leq M s^\star\}$). For each piece, we apply Lemma \ref{apx-test}. Then we obtain that
\begin{align*}
&\hspace{-1cm}\sup_{\|\Sigma - \Sigma^\star\| > M_2 \epsilon_n} \mathbb{E}_{f} (1-\phi_n)\\
& \leq \exp\left(
C_3 s - \frac{C_3 M_2 n\epsilon_n^2}{4} 
\max\left\{
1, 
\frac{M_2}{(\sqrt{M_2} + 2)^2 \|\Sigma^\star\|^2}
\right\}
\right)\\
& \leq 
\exp\left(
C_3 s - \frac{C_3M_2 n \epsilon_n^2}{4}
\right).
\end{align*}
Finally, we sum up the above small pieces and obtain that
\begin{align*}
 \sup_{f \in \mathcal{F}_n: \|\Sigma - \Sigma^\star\| \geq M_2 n \epsilon_n^2}
 \mathbb{E}_f(1 - \phi_n)
& \leq 
 \sum_{|S| \leq M s^\star} 
 \exp\left(
C_3 s - \frac{C_3 M_2 n \epsilon_n^2}{4}
\right)\\
& \leq 
\exp\left( \log(M s^\star) + 
C_3 M s^\star - \frac{C_3 M_2 n \epsilon_n^2}{4}
\right) \\
& \lesssim 
\exp(- C_3' n \epsilon_n^2/4),
\end{align*}
where $C_3' > C_3(M_2/4 - M - \log M/C_3)$.

So far, we have verified all the three conditions. Therefore, we obtain that 
\begin{align*}
    & \mathbb{E}_{f^\star} \Pi\Big(\|\Sigma - \Sigma^\star\| > M_2 \epsilon_n, 
    |S| \leq M_2 s^\star | X\Big)\\
    & \quad \leq 
    \mathbb{E}_{f^\star} 
    \Bigg[\Pi\Big(\|\Sigma - \Sigma^\star\| \geq M_2 \epsilon_n,
    |S| \leq M_2 s^\star | X\Big)  \\
    & \hspace{2cm} \quad \times \Pi\left( \int f/f^\star d\Pi(\theta) \leq \exp(-C_1n\epsilon_n^2)\right)(1-\phi_n)
    \Bigg]\\ 
    & \quad \quad 
    + \mathbb{E}_{f^\star} \phi_n
    + \mathbb{E}_{f^\star} 
    \Pi \left( 
    \int f/f^\star d\Pi(\theta) \geq \exp(-C_1n\epsilon_n^2)
    \right)\\
    & \quad \lesssim
    \delta_n \to 0,
\end{align*}
where $  \delta_n = \exp(- (C_1 \vee C_2) n\epsilon_n^2 )
    + \exp \left(
    C_3 M_2 s^\star - C_3 M_2^2 n \epsilon_n^2/(4 b_5^2)
    \right)$.
Thus the proof of (\ref{Thm-1:eqn-1}) is complete.    
    
To prove (\ref{Thm-1:eqn-2}), by David-Kahn $\sin \theta$ theorem (see Theorem \ref{apx-davis-kahan}),
$$\big\{\|\Sigma - \Sigma^\star\| \leq M \epsilon_n\big\} \subset \big\{\|UU' - U^\star {U^\star}'\| \leq M' \epsilon_n\big\},$$ where $M' = M/b_5$, thus (\ref{Thm-1:eqn-1}) implies (\ref{Thm-1:eqn-2}).
\end{proof}

\subsection{Proof of Theorem \ref{Thm: variational-post-contr-rate}}

\begin{proof}
We apply Theorem \ref{Thm: general-theory-vp}.
Let $\rho = 2$ and $L = \|\Sigma - \Sigma^\star\|$, we only need to verify the following conditions:
\begin{eqnarray}
& \Pi(D_2(f^\star, f) \leq n\epsilon_n^2) \geq \exp(-C_4n\epsilon_n^2), &
\label{proof-mfvb-cond1}\\
&  - \log P(\theta) \leq C_5n\epsilon_n^2, &
\label{proof-mfvb-cond2}
\end{eqnarray}
where  $D_2(f^\star, f)$ is the $\chi^2$-divergence between $f^\star$ and $f$, and 
$$\Theta \subset \Big\{
\theta: K(f^\star, f) \leq C_1n\epsilon_n^2, \
\log \left(dQ(\theta)/d\Pi(\theta)\right) \leq C_2 n\epsilon_n^2
\Big\}.$$

We first verify (\ref{proof-mfvb-cond1}).
Recalling that $f$ and $f^\star$ are both independent multivariate normal densities, we obtain
\begin{align*}
    D_2(f^\star, f) & 
    =\frac{1}{2} \log\left(
    \int {(f^\star)^2}/{f} \right)\\
    &= -\frac{n}{4}
    \left(
        \log \det(\widetilde \Sigma) + 
        \log \det \left(2I_p - \widetilde \Sigma\right)
    \right)\\
    & = 
    -\frac{n}{4} \sum_{j=1}^p
    \Big[
        \log (1+(\rho_j-1))  + \log (1-(\rho_j - 1))
    \Big] \\
    & \leq n \sum_{j=1}^p (\rho_j-1)^2
\end{align*}
where $\widetilde\Sigma = \Sigma^{-1/2}\Sigma^\star \Sigma^{-1/2}$
and $\rho_j$ is the $j$-th largest eigenvalue of $\widetilde\Sigma$.
The upper bound in the last display is obtained by first applying Taylor expansion to the two log functions and then using the inequality 
$x^2 + x^4/2 + x^6/3 + \cdots \leq 4x^2$ when $|x| < 1/2$.
The last display leads to the following inequalities
\begin{align*}
\Pi\big(D_2(f^\star, f) \leq n\epsilon_n^2\big) 
\geq \Pi\big(\|\widetilde \Sigma - I_p\|_F \leq \epsilon_n\big)
\geq \Pi\big(\|\Sigma - \Sigma^\star\|_F \leq \sigma^2 \epsilon_n\big).
\end{align*}
Similar to the proof of Lemma \ref{lemma-1}, the probability at the right hand side of the above inequality is bounded below by
$$\Pi(\|\beta - \beta^\star\|_F \leq d_n' \big| r=r^\star) \Pi(r^\star = r),$$
where $d_n' = \sigma^2\epsilon_n/\|\beta^\star\|_{q,1}$.
Under Assumption \ref{assumps}, we obtain $\Pi(r^\star = r) \geq \exp(-a_3 r^\star)$.
The first probability in the product, it is bounded in
(\ref{pf-post-contr-rate-1.4}) with $d_n$ replaced with $d_n'$.
We then obtained (\ref{proof-mfvb-cond1}) by applying the same argument as in (\ref{pf-post-contr-rate-1.4}).

To verify (\ref{proof-mfvb-cond2}),
we choose $\widetilde{P} = \prod_{j=1}^p \widetilde{P}_j$, where $$\widetilde{P}_j = \gamma_j^\star \mathcal{N}(0, I_{r^\star}/ \lambda_1) + (1-\gamma_j^\star) \delta_0.$$
Clearly, $\widetilde{P}_j \in Q^{MF}$. 
Then (\ref{proof-mfvb-cond2}) can be written as
\begin{align}
\label{proof-mfvb-cond3}
    \widetilde{Q}
    \left(
        KL(f^\star, f) \lesssim n\epsilon_n^2, \
        \log \frac{d\widetilde{Q}(\theta)}{d\Pi(\theta)} \lesssim n\epsilon_n^2
    \right) \geq \exp(-C_5 n \epsilon_n^2)
\end{align}
Apply a similar argument as that in the proof of Theorem \ref{Thm: Post-cont-rate} yields
\begin{align}
\label{mf-posterior-1}
   &\hspace{-1cm} \widetilde{Q}
    \left(
        KL(f^\star, f) \lesssim n\epsilon_n^2, \
        \log \frac{d\widetilde{Q}(\theta)}{d\Pi(\theta)}
        \lesssim n\epsilon_n^2
    \right)\\
    &\geq 
    \widetilde{Q}
        \left(
        \sum_{j \in S^\star} \|\beta_j - \beta_j^\star\|_2 \leq d_n', \
        \log \frac{d\widetilde{Q}(\theta)}{d\Pi(\theta)}
        \lesssim n\epsilon_n^2
    \right).
\end{align}
Note that
\begin{align}
\label{bound-1}
& \log d\widetilde{Q}(\theta) - \log d\Pi(\theta) \nonumber \\
& \quad \leq \log d\widetilde{Q}(\theta) - \log d\Pi(\theta|S = S^\star) -
\log \left(\frac{\pi(|S^\star|)}{{p \choose s^\star}}\right).
\end{align}
Plugging in the density function of the multivariate normal distribution and the expression of $g(\theta)$, and then applying Jensen's inequality
$$\log \int g(\theta|A) d\Pi(A) \geq \int \log (g(\theta|A) \pi(A)) dA,$$ we obtain 
\begin{align}
    & \log d\widetilde{Q}(\theta) - \log d\Pi(\theta|S = S^\star) 
    \log (\pi(|S^\star|)) + \log {p \choose s^\star}
    \nonumber \\
    & \quad \leq c_1 r^\star s^\star \log (r^\star)  + c_2s^\star \log p - \lambda_1\sum_{j\in S^\star} \|\theta_j - \theta_j^\star\|_2^2 + \lambda_1 \sum_{j \in S^\star} \|\beta_j \|_q^m \nonumber \\
    & \quad \leq 
    c_3 r^\star s^\star \log (r^\star) + \lambda_1 \sum_{j \in S^\star} \|\beta_j - \beta_j^\star\|_1 +
    \sum_{j \in S^\star} \|\beta_j^\star\|_1\nonumber\\
    & \quad \leq c_4 r^\star s^\star \log p + \lambda_1 \sum_{j \in S^\star} \|\beta_j - \beta_j^\star\|_1,
        \label{bound-2}
\end{align}
where $c_1, \dots, c_4$ are positive constants. The first inequality in the last display is by Assumption \ref{assumps} and the upper bound $\log {p \choose s^\star} \lesssim s^\star \log p$. The second inequality is obtained using the triangle inequality that $\|a \|_q^m \leq \|a - b\|_q^m + \|b\|_q^m$.
The last inequality is obtained by plugging-in the upper bound of $\sum_{j \in S^\star} \|\theta^\star\|_1$ in Assumption \ref{assump:truevalue}, as $\|\beta^\star_j\|_1 \leq \|\theta^\star_j\|_1$ for all $j$.
Let $d_n'' = \sigma^2 \lambda_1/n\epsilon_n$, what left to show is the following:
\begin{align*}
&\hspace{-1cm}\widetilde{Q}
\Bigg(
\sum_{j \in S^\star} \|\beta - \beta^\star\|_1 \leq d_n''
\Bigg)\\
& \geq \widetilde{Q} \Bigg(
	\sum_{j \in S^\star} \|\beta_j - \beta_j^\star\|_2 \leq d_n'/\sqrt{r^\star}
\Bigg) \\
& \geq 
\exp \Bigg(-\lambda_1 \sum_{j \in S^\star} \|\theta_j^\star\|_2 \Bigg)
\left(1 - 
\Phi\Bigg(
|x| \geq \frac{d_n''}{\sqrt{{r^\star}^3}s^\star \lambda_1}\Bigg) \right)^{s^\star r^\star}\\
& \geq 
\exp(- n\epsilon_n^2) 
\exp\left(r^\star s^\star 
\log ({d_n''}^2 / (2{r^\star}^3 {s^\star}^2 \lambda_1^2)\right)\\
& \geq 
\exp\left(- n\epsilon_n^2 - 2s^\star r^\star \log n - s^\star r^\star 
\log (2{r^\star}^3 s^\star/\sigma^4)\right) \\ 
& \geq \exp(-5n\epsilon_n^2),
\end{align*}
where the notation $\Phi$ stands for the cumulative distribution function  of the standard normal distribution.
Here, the second inequality is obtained by changing the variables from $\beta_j-\beta^\star_j$ to $\check{\beta_j}$ and use that $\|\beta\| = \|\theta\|$.
The fourth inequality is obtained by using the inequality 
$\lambda_1 \|\theta^\star\|_{1,1} \leq n\epsilon^2_n$.
The last inequality uses $r^\star \leq \log p/\log n$ in Assumption \ref{assump:truevalue}
and $s^\star \ll n$.
\end{proof}

\subsection{Proofs of Lemma \ref{lemma-convergence-EM}}
\begin{proof}
We apply the parameter expansion twice, but we only prove the result for the first parameter expansion, since the proof for the second parameter expansion is similar. 

The expanded parameter in chosen to be $A$, an orthogonal matrix.
The EM defines a map $P$ such that $\Delta^{(t+1)} = P\Delta^{(t)}$ at the $t$-th iteration. Then, by Taylor's theorem, 
$\Delta^{(t+1)} - \Delta^\star \approx DP (\Delta^{(t)} - \Delta^\star)$, where $DP$ is the derivative of $P$ evaluated at $\Delta^\star$ and the speed of convergence is governed by the largest eigenvalue of the matrix $D\Psi$.
Define $S = I - DP$ and $S = I^{-1}_{com}(\Delta) I_{obs}(\Delta)$.
After expanding the parameter, we obtain 
\begin{align*}
I_{obs}(\widetilde \Delta)
=
\begin{pmatrix}
I_{obs}(\Delta) & 0\\
0 & 0
\end{pmatrix}
\quad \text{and}\quad
I_{com}(\widetilde \Delta) 
=
\begin{pmatrix}
I_{com}(\Delta) & F\\
F' & G
\end{pmatrix},
\end{align*}
where 
$F = -\frac{\partial^2 Q(\widetilde \Delta| \widetilde \Delta)}{\partial \Delta \partial \Psi'}$
and $G = -\frac{\partial^2 Q(\widetilde \Delta| \widetilde \Delta)}{\partial \Psi \partial \Psi'}$.

We now calculate $F$ and $G$. Assuming $m = 1$ and $\sigma^2 = 1$, the objective function is
\begin{align}
Q &= C - \sum_{j=1}^p \Big(\beta_j M \beta_j'/2 - \beta_j M_L d_j + pen_j  \|\beta_j\|_q \Big) + f(\widetilde{\bgamma}, \kappa)
\nonumber
\\
& = C - \sum_{j=1}^p \Big(\theta_j M \theta_j'/2 - \theta_j A M_L d_j + pen_j \|\theta_j A\|_q\Big) + f(\widetilde{\bgamma}, \kappa),
\end{align}
where $pen_j = \widetilde \gamma_j \lambda_1 + (1-\widetilde \gamma_j) \lambda_0$ and $$f(\widetilde\gamma, \kappa)
= (\|\widetilde{\bgamma}\|_1 + \alpha_1 - 1) \log \kappa + (p - \|\widetilde{\bgamma}\|_1+\alpha_2- 1) \log (1-\kappa).$$
If $q = 2$, then $\|\beta_j\|_2 = \|\theta_j\|_2$, we obtain $G = 0^{r\times r}$ and 
$F = -\sum_{j=1}^p M_L d_j$.
Denote
\begin{align*}
I^{-1}_{com}(\widetilde \Delta) 
=
\begin{pmatrix}
V_{\Delta, \Delta} & V_{\Delta, m}\\
V_{\Delta, m}' & V_{m,m}
\end{pmatrix},
\end{align*}
we have
\begin{align*}
V_{\Delta, \Delta} 
& = I_{com}^{-1}(\Delta) + I_{com}^{-1}(\Delta) F(G - F' I_{com}^{-1}(\Delta) F)^{-1} F'I_{com}^{-1}(\Delta) \\
& =  I_{com}^{-1/2}(\Delta) 
\left[
I - I_{com}^{-1/2}(\Delta) F
(F' I_{com}^{-1/2}(\Delta) F)^{-1} F' I_{com}^{-1/2}(\Delta)\right]
I_{com}^{-1/2}(\Delta)\\
& < I_{com}^{-1}(\Delta).
\end{align*}

If $q = 1$, then $G = 0^{r\times r}$ and $F = - \sum_{j=1}^p M_L d_j + pen_j \times \text{sign}(\beta_j)$.
We can also obtain that $V_{\Delta, \Delta} < V_{com}^{-1}(\Delta)$. 
Therefore, the smallest eigenvalue of $S$ decreases after the parameter expansion and hence, the speed of the PX-EM algorithm increases, as the largest eigenvalue $D\Psi$ increases. 
\end{proof}

\section{Auxiliary lemmas}
\label{sec:aux}

\begin{lemma}
\label{gammaUpperBound}
For $X \sim \text{Gamma}(r, \lambda)$, then $P(X> b) \leq \exp((r - \lambda b)/2).$
\end{lemma}
\begin{proof}
From page 29 of \citet{bouc13}, we have
\begin{align*}
    P(X > b) \leq 
    \exp\left(
        -r
        \left( 
            1 + \lambda b / r - \sqrt{1 + 2\lambda b / r}
        \right)
    \right).
\end{align*}
Now use the inequality $1 + b - \sqrt{1+2b} \geq (b-1)/2$ for $b > 0$, we obtain the desired upper bound.
\end{proof}

\begin{lemma}[\citet{gao15}]
\label{apx-test}
For random i.i.d. variables $Y_i \sim \mathcal{N}(0, \Gamma^\star)$, $i = 1, \dots, n$, $Y_i \in \mathbb{R}^d$, $d < n$, and $C_3, M > 0$, there exists a test function $\phi$ such that 
$$P_{\Sigma^\star} \phi(Y^n) 
    \leq \exp\left(C_3 d - \frac{C_3M^2 n\epsilon_n^2}{4 \|\Sigma^\star\|^2}\right)
    + 2\exp(C_3d - C_3M^{1/2}n),$$
 \begin{align*}
 	&\hspace{-1cm} \sup_{\{\Gamma: \|\Gamma - \Gamma^\star\| > M\epsilon\}}
 	P_{\Sigma} (1 - \phi(Y^n))\\
 	&\leq \exp
 	\left(
 	C_3d - \frac{C_3 Mn\epsilon_n^2}{4}
 	\max\left\{
 	1, \frac{M}{(\sqrt{M} + 2)^2 \|\Gamma^\star\|^2}
 	\right\}
 	\right).
 \end{align*}
\begin{proof}
See the proof of Lemma 5.7 in \citet{gao15}.
\end{proof}
\end{lemma}

\begin{theorem}[The Davis-Kahan $\sin \theta$ theorem]
\label{apx-davis-kahan}
Let $\Sigma$ and $\widehat\Sigma$ be two symmetric matrices with eigenvalues $\lambda_1 \geq \cdots \geq \lambda_p$ and $\widehat\lambda_1 \geq \cdots \geq \widehat\lambda_p$. Fix $1 \leq r \leq s \leq p$ and let $d = s-r+1$,
and let $V$ be first $d$ eigenvectors of $\Sigma$ and use a similar definition for $\widehat V$. Let $\delta = \inf\{|\widehat\lambda - \lambda|: \lambda \in [\lambda_s, \lambda_r]\}$ and assume that $\delta > 0$,
then 
\begin{align*}
    \|\widehat V\widehat V' - VV' \|_F  \leq \delta^{-1} \|\widehat\Sigma - \Sigma\|_F\quad\text{and}\quad
    \|\widehat V\widehat V' - VV' \| \leq \delta^{-1} \|\widehat\Sigma - \Sigma\|.
\end{align*}
\end{theorem}

\begin{theorem}[Theorems 2.1 and 2.4 of \citet{zhang20}]
\label{Thm: general-theory-vp}
Suppose that for a sequence of $\epsilon_n$ with $\epsilon_n \to 0$ and $n\epsilon_n^2 \to \infty$.
Consider a loss function $L$ such that $L(f^\star, f) \geq 0$ and let $D_\rho$ is the $\rho$-R\'enyi divergence, $\rho > 1$, let $\Theta \in \mathcal{F}_n$, the sieve, and $\phi_n$ be a sequence of test function, if 
\begin{align}
    & \Pi(D_\rho(f^\star, f) \leq C_1n\epsilon_n^2) \geq \exp(-C_2 n \epsilon_n^2),
    \label{mf-cont-rate-cond1}\\
    & \Pi(\mathcal{F}_n^c) \leq \exp(-Cn\epsilon_n^2), 
    \label{mf-cont-rate-cond2}\\
    & \mathbb{E}_{f^\star}(\phi_n) + \sup_{\theta \in {\mathcal{F}_n \cap \{L(f^\star, f) \geq C_3n\epsilon_n^2\}}} \mathbb{E}_f(1-\phi_n) \leq \exp(-Cn \epsilon^2),
    \label{mf-cont-rate-cond3}
\end{align}
then for the variational posterior $\widehat q$, we have
\begin{align*}
    \mathbb{E}_{f^\star} \widehat Q (L(f, f^\star)) \leq M n(\epsilon_n^2 + \gamma_n^2),
\end{align*}
for some constants $C_1, C_2, C_3$, and $M$ and $C > C_1+C_2+2$,
where 
\begin{align}
\gamma_n^2 & = \frac{1}{n} \inf_{q \in Q^{MF}} \mathbb{E}_{f^\star} K(q, \pi(\beta|X))
\leq \inf_{q \in Q^{MF}} R(q),
\end{align}
where $R(q) = \frac{1}{n}\left(K(q, \pi(\beta|X) + \mathbb{E}_{q}[K(f^\star, f)]\right)$.

Furthermore, if $q \in Q^{MF}$, denote a subset $\Theta = \prod_{j=1}^p \Theta_j$ such that
\begin{align*}
    \Theta \subset 
    \left\{
        \beta: KL(f^\star, f) \leq C_1 n\epsilon_n^2, \
        \log \frac{dQ(\beta)}{d\Pi(\beta)} \leq C_2 n\epsilon_n^2
    \right\}, 
\end{align*}
and 
\begin{align}
\label{mf-cont-rate-cond4}
    - \sum_{j=1}^p \log q_j(\Theta_j) \leq C_3n\epsilon_n^2,
\end{align}
then for some positive constants $C_1, C_2$, and $C_3$, we have 
\begin{align*}
    \inf_{q \in Q^{MF}} R(q)\leq (C_1+C_2+C_3) \epsilon_n^2.
\end{align*}
\end{theorem}

\bibliographystyle{chicago}
\bibliography{citation.bib}

\makeatletter\@input{supp.tex}\makeatother